# Part II: Technical Options for Flexible Hardware-Enabled Guarantees

James Petrie and Onni Aarne
*April 2025*

## About This Report

This is the second part of a three-part exploration of flexible hardware-enabled guarantees (flexHEGs), commissioned by ARIA. This document can serve as a stand-alone discussion of flexHEGs for technical audiences. Readers who are interested in additional discussion of the motivation of the proposal should read Part I, and readers who are interested in uses of flexHEG for international treaties about AI development should read Part III.

## Table of Contents

# Introduction

Frontier AI models could pose serious risks to public safety and international security [1]. Because of these potential risks, it may soon be important for frontier AI developers to be able to make credible guarantees about what AI development they are or are not doing. It would be very useful if AI developers could provide these guarantees without having to reveal proprietary information about how their models are developed.

Frontier AI development depends heavily on specialized hardware - particularly AI accelerators running in massive data centers. These accelerators are a natural node for structuring guarantees about AI development, as a large number of accelerators are needed and their supply chain is highly concentrated [2]. In some contexts, such as domestic regulation, usage of these accelerators could be verified by trusted intermediaries like cloud providers [3]. However, for international agreements about AI development, there may be no entity that all parties would trust with such deep visibility into their AI programs.

To address this challenge, we propose Flexible Hardware-Enabled Guarantees (flexHEG) - a system that uses open source and privacy-preserving hardware to enable AI developers to make verifiable claims about their compute usage. The system would be:

- Privacy-preserving: Developers could verify compliance with guarantees without revealing sensitive details about their work.
- Flexible: The verification capabilities could be updated over time as technology and governance needs evolve.
- Trustworthy: The system would be open source and auditable to provide confidence that it does not contain backdoors.
- Secure: Physical and cryptographic protections would prevent circumvention of the guarantees.

FlexHEG is composed of two main components: an auditable Guarantee Processor that monitors accelerator usage and verifies compliance with specified rules, and a Secure Enclosure that protects against physical tampering. Because both AI technology and governance needs are rapidly evolving, the system is designed to flexibly support many types of guarantees through updateable verification capabilities. The Guarantee Processor could be configured to only accept updates approved by appropriate stakeholders - the implications of this for international governance are analyzed in Part III of this report series.

For verification use cases, it is sufficient for the flexHEG Secure Enclosure to be tamper-evident, ensuring that tampering with a large number of devices could not go undetected. For use cases that require guarantees about future accelerator usage, the secure enclosure should also be able to trigger a tamper response that permanently disables the accelerator before an attacker can circumvent the guarantee logic or extract secret keys.

This report explores the technical options for implementing flexHEG, with a focus on approaches that could be deployed relatively quickly. First, we establish concrete design requirements, arguing that it should be auditable and highly secure to be most useful for emerging AI governance needs (FlexHEG Design Goals). We then review modern AI data centers and AI development workflows that flexHEG must integrate with (Background on Frontier AI Development).

Next, we evaluate different strategies for modifying AI accelerators, from software changes to custom hardware (Potential Accelerator Modifications). We then analyze specific design choices for the two main components - the Guarantee Processor and Secure Enclosure - discussing tradeoffs between security and deployment speed (Hardware Components).

For more robust guarantees, we propose a specific flexHEG architecture that gives the Guarantee Processor direct access to the accelerator's data path (Interlock-Based Design). This "Interlock" design enables robust verification without requiring trust in the accelerator's internal operation. Modifying the accelerator's Network Interface Controller (NIC) is a promising strategy for implementing this design that may be achievable by third parties.

Next, we describe concrete approaches to implementing guarantees, from basic auditing to sophisticated automated verification ([General-Purpose Guarantees](#)). A detailed example of verifiable FLOP counting for multi-accelerator training demonstrates how these mechanisms could work in practice.

Finally, we outline a roadmap that a major third-party R&D effort could take to develop flexHEG ([Future Work](#)). This development could be sped up and directly integrated into existing accelerators with the help of accelerator manufacturers.

# FlexHEG Design Goals

To provide the capabilities required for the use cases discussed in [Part I](#) and [Part III](#) of this report series, a full flexHEG system has the following properties (although systems that meet only some of the high-level goals are also useful in more specific contexts) :

- Enables AI developers to make guarantees about past and future workloads performed by their accelerators
- Is secure against attempts to circumvent guarantees
- Flexibly allows guarantees to be updated if authorized by the appropriate parties
- Is trusted not to contain backdoors
- Integration doesn't significantly burden model developers
- Can be deployed by 2027 to be included in major data center build outs

To ensure that a given future flexHEG implementation meets these high-level goals, we specify several lower-level requirements, which are summarized in Table 1 and motivated below. The table entry indices are referenced with brackets throughout the text in this section.

| Category | Design Goals |
|---|---|
| Enables AI developers to make guarantees about past and future workloads | 1. Can execute general-purpose logic <br> 2. Access to accelerator workload data <br> 3. Accurate timekeeping <br> 4. Ability to control accelerator usage <br> 5. Automated guarantee checking |
| Is secure against attempts to circumvent guarantees or exfiltrate data | 6. Protection against non-invasive attacks <br> 7. Resistance to invasive attacks <br> 8. Confidential and authenticated communication <br> 9. Confidential data storage |
| Allows guarantees to be updated if authorized | 10. Secure update process |
| Is trusted not to contain backdoors | 11. No surveillance/control backdoors <br> 12. No intentional security flaws <br> 13. Third-party verifiability of system integrity |
| Integration doesn't significantly burden model developers | 14. Minimal additional cost <br> 15. Minimal additional power consumption <br> 16. Compatibility with cooling solutions <br> 17. Standard rack/interconnect compatibility <br> 18. Minimal impact on interconnect latency/throughput <br> 19. Infrequent false positives <br> 20. Minimal impact on hardware reliability <br> 21. Simple to demonstrate compliance with guarantees <br> 22. Minimal impact on development workflow |
| Can be deployed by 2027 to be included in major data center build outs | 23. Ready for large-scale deployment by 2027 <br> 24. Compatibility with different types of accelerators <br> 25. Short notice integration |

Table 1: FlexHEG design goals, separated by category.

**Enables AI developers to make guarantees about past and future workloads performed by their accelerators**

The system is able to execute general purpose logic (1) so that it can flexibly perform a variety of guarantees. The flexHEG system has access to accelerator workload data (2) so that it has the necessary information to check guarantees; and accurate timekeeping (3) so that it can support time-based guarantees. The flexHEG system can enable control of accelerator usage (4) so that it can nondestructively enforce that guarantees are adhered to, which is necessary so that a developer can credibly make guarantees about future usage of accelerators. Guarantees can be checked in an automated way (5) so that guarantee checks can be performed in a sufficiently short time, and model details do not need to be shared with auditors (i.e. a hardware-backed zero-knowledge proof).

**Is secure against attempts to circumvent guarantees or exfiltrate data**

The flexHEG system is secure against non-invasive attacks (6) so that guarantees cannot be circumvented with scalable attacks. The system is resistant against invasive attacks (7), with either tamper evidence to disincentivize covert adversaries, or a tamper responsive enclosure that can trigger permanent accelerator disablement to disincentivize overt adversaries. Defense against overt adversaries is required so that guarantees about future usage are a credible commitment that cannot be skipped later by physically disabling the guarantee system. A more detailed exploration of the type of attacks we hope to defend against is provided in [Appendix A: Threat Models](#).

Communication between flexHEG systems is confidential and authenticated (8) to prevent network snooping being used to circumvent guarantees about how data is shared, and so that multiple flexHEG systems can coordinate around guarantees which pertain to many-accelerator workloads. Additionally, accelerator and flexHEG data at rest can be stored in a confidential way (9) so that guarantees cannot be circumvented by an adversary obtaining data with e.g. scanning equipment, and claims from a flexHEG system cannot be falsified using a stolen key.

**Allows guarantees to be updated if authorized by the appropriate parties**

FlexHEG guarantees and the logic to check them can be updated if and only if the update has been authorized by a sufficient number of pre-specified stakeholders (10). Different types of updates could require approval from different stakeholders, and could depend on other conditions being met. Updates could potentially modify the rules for accepting future updates.

**Is trusted not to contain backdoors**

FlexHEG must be trusted to not contain backdoors for secret surveillance or control (11), or intentional security flaws (12) that would allow insiders to circumvent guarantees. Additionally, it is possible for third parties to verify that the flexHEG system is operating as expected (13).

**Integration doesn't significantly[1] burden model developers**

Integration of flexHEG capabilities should not create significant burdens for AI developers. The addition of flexHEG hardware should not significantly increase the total datacenter cost (14). Additional power consumption should be minimal (15). The system is ideally compatible with both air and liquid cooling solutions without significantly increasing the thermal resistance (16). The flexHEG form factor should fit within common server configurations and maintain compatibility with existing interconnects (17). Interconnect latency and throughput should not be significantly impacted for relevant workloads (18). False positives in tamper detection or guarantee checks should be rare (19) to avoid disruptions. Accelerator reliability should not be substantially reduced (20), either by allowing regular maintenance, or by modifying configurations to tolerate partial hardware failure. Demonstrating compliance with guarantees should not be overly time consuming (21). Finally, flexHEG should not significantly impede development or debugging workflows (22).

**Can be deployed by 2027 to be included in major data center build outs**

Massive data center build outs are planned between 2025-2030 [4], [5], and AI development past 2030 may occur on the same hardware and infrastructure, even if transformative AI is several years later. In order to reach its full potential, flexHEG systems will likely need to be deployed at scale by 2027 or sooner (23). Additionally flexHEG should be compatible with the different types of accelerators (24) that are used by frontier AI developers, ideally on short notice (25) for cases where countries were not able to collaborate years in advance of a treaty.

# Potential Accelerator Modifications

In this section, we discuss four ways that AI accelerators could be modified to meet the flexHEG design goals, with tradeoffs between security, time to readiness, and sophistication of guarantees. For more background on frontier AI workloads and the data center hardware they run on, refer to Appendix A for a brief summary.

Accelerators could be modified to include flexHEG capabilities using integrated flexHEG components, retrofitted flexHEG components, firmware modifications, or software modifications. Including flexHEG directly in the accelerator die or PCB would provide the greatest design flexibility but would take the longest to deploy. To speed up deployment, flexHEG components could be separately manufactured and retrofitted to existing accelerators. Finally, firmware and software changes are the fastest to deploy, but rely on the limited hardware security that is already on devices. The advantages and disadvantages of these options are discussed below.

---

[1] What counts as a "significant" burden depends on the context, but as a rough guess this might be a 1-2% percent increase in cost in typical scenarios, 5% if needed to satisfy regulations, and 20% if there is high-level buy-in.

## Hardware Modifications (Integrated)

Specialized flexHEG hardware could be integrated directly into the accelerator PCB or main chip, which would provide the best security and data access. If the Guarantee Processor is on the main chip, it would be much more difficult for attackers to modify signals that stay within the chip. Additionally, supply chain attacks that divert a fraction of components before adding a Guarantee Processor would be more difficult to execute (because the Guarantee Processor would be on the same mask). A purpose-built Guarantee Processor could be designed to prioritize security and auditability, as discussed in the Guarantee Processor section.

The Secure Enclosure could also be added at the chip level, which may improve security by reducing the size of the attack surface (however, it may still be important to also have a larger secure enclosure that prevents direct access of tools to the chip surface).

If flexHEG hardware were deeply integrated throughout an accelerator, it may be more difficult for an attacker to physically disable it. However, deep integration would make external audits challenging and reduce portability to other accelerator types. A potential solution is to have dedicated blocks within the hardware for the Guarantee Processor that can be open sourced for external audits.

There are three operational downsides to integrated hardware changes:
- They require significant assistance from the accelerator manufacturer, which could cause challenges due to, among others, proprietary IP.
- They likely require years of preparation to add to each type of accelerator (among others due to multi-year-long manufacturing cycles).
- They cannot be added to existing accelerators, which is a downside both for the absolute year they could be available, and for last-minute integration into novel accelerator types.

The lag time for hardware- integrated flexHEG depends on the time for it to be designed and manufactured, and the time for older compute to be displaced. An investigation by Epoch AI, a research institute focused on analyzing key trends in AI, suggests that, counting from the time a new hardware generation is launched, it would take 2.7 - 3.9 years for the old accelerator generation to no longer be useful for frontier development (see Figure 2) [6]. If the current trend for deployed compute slowed[2], this estimate would increase (because it would take longer for prior compute to be displaced).  Additionally, as a rough estimate[3] it might take 1-4 years to add changes to all leading accelerator designs. Overall, this estimate predicts 3.7 - 7.9 years for flexHEG with integrated hardware to displace non-flexHEG accelerators in frontier AI development (starting when the accelerator manufacturer begins actively working on it).

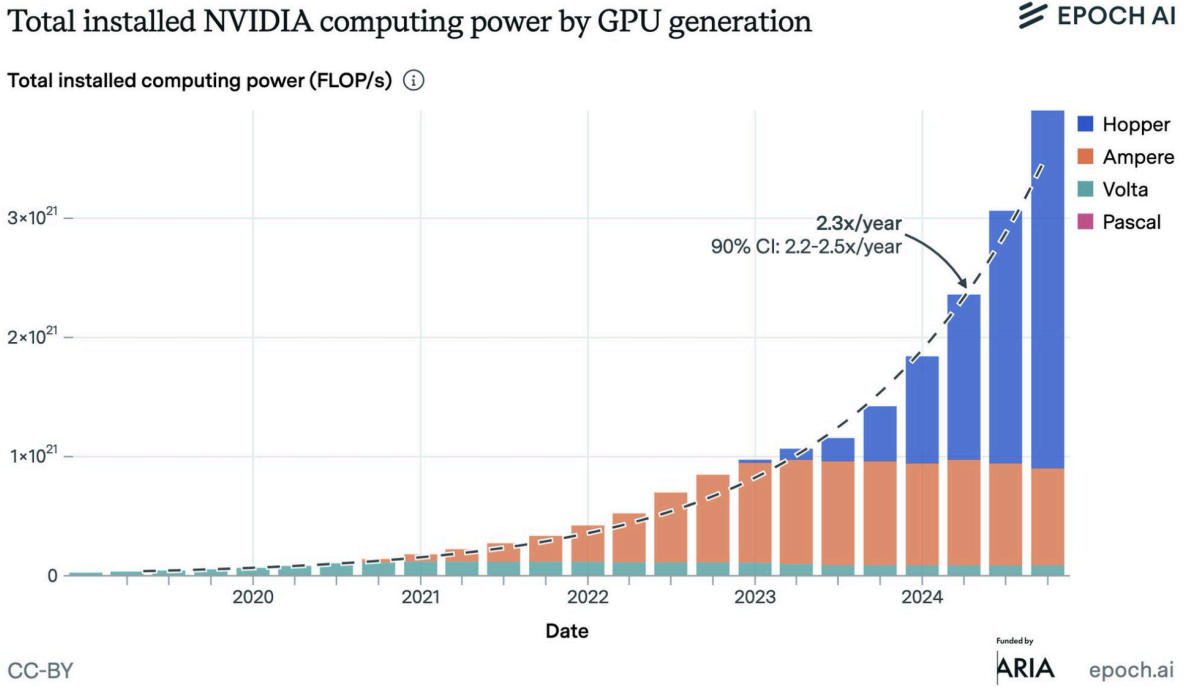

Figure 2: Installed NVIDIA computing power by accelerator generation over time. Source: Epoch AI [6].

## Hardware Modifications (Retrofitted)

If the modifications can instead be retrofitted to existing accelerators, the time to deployment and reliance on accelerator designers can be reduced. Retrofittability may be especially important if flexHEG needs to be quickly added to accelerators used by countries that are not allowed to purchase the newest generation of NVIDIA chips. Because retrofitted hardware is added after manufacturing, additional supply chain security may be needed. Compared to integrated hardware, retrofittable hardware may be easier for an attacker to decouple from the accelerator. This can potentially be defended against with a retrofittable Secure Enclosure, which is discussed in more detail in Secure Enclosure.

---

[2]  It's unclear whether this exponential growth in compute will continue past 2030 because it would start to reach a significant fraction of electricity production [7] (and predictions for AI trajectories more than five years in the future are difficult to make confidently).
[3] Nvidia does not publicize their engineering design timelines, but some sources say major architectural modifications take 3 years from design to rollout.

One option to retrofit a Guarantee Processor is to add it as an additional component that can make external measurements, as done with the power-measurement prototype. To get additional data access, the Guarantee Processor could potentially be plugged into an open PCIe slot (for example, NVL72 trays have 8 slots for PCIe drives, which may not all be used [8]). Nvidia supports third party direct memory access to HBM [9], so a third-party device could potentially be set up to read from HBM[4].

Similarly, an existing retrofittable component on the accelerator could be replaced with a flexHEG version. For example the Network Interface Controller, or Hardware Management Controller (also called BMC) could be replaced to function as the Guarantee processor, as discussed in Guarantee Processor.

## Firmware Modifications

Firmware is code that runs on computing devices and is generally responsible for lower-level device functioning. Several components within an accelerator will typically have their own firmware. A security advantage of firmware over software is that devices that have Secure Boot [10] can only run firmware that has been cryptographically signed by the manufacturer (as opposed to software, which can be easily modified or replaced). Devices that also have Rollback Protection [10] can only run firmware binaries that have a version number that is greater than a value stored in secure memory. Additionally, because firmware can run directly on the AI chip, it has the same security benefit as integrated hardware that signals are more difficult to tamper with.

There are several components on existing accelerators that could potentially be reprogrammed with a firmware update to support flexHEG logic. Components used for compute, networking, and memory could potentially be repurposed, with different tradeoffs between feasibility, security, and the types of guarantees that could be supported. These options are discussed in more detail in the Guarantee Processor section. One downside is that it may be more difficult (compared to using a dedicated Guarantee Processor) for auditors to verify that firmware on a proprietary chip is functioning as intended.

As a relatively simple example, it may be possible to create modified firmware for existing accelerators that would require occasional authorization from the manufacturer or a regulator in order for the accelerator to continue operating [11], [12], [13]. By combining this with verification software (discussed in next subsection), it could be possible to structure guarantees about future usage without making hardware changes. I.e., receiving updated licenses permitting accelerator usage could be contingent on auditors having received evidence that ongoing accelerator usage met the guarantees.

---

[4] Although there are some security risks with this because a sophisticated actor may be able to spoof PCIe reads that have to physically route through multiple components. As another potential limitation, the current implementation of confidential computing blocks direct memory access from third party devices, which may have to be modified to support this use case (this might have to be done by Nvidia anyways to support efficient multi-node confidential computing).

While firmware with Secure Boot can be more challenging to circumvent than software, it is not completely resistant to physical attacks. The most common of these attacks is to modify or replace the firmware, then apply a voltage glitch at the exact time during Secure Boot when the cryptographic signature is being checked[5] [14]. For devices without protections against voltage glitching, this can cause the signature check to incorrectly pass, allowing the use of unauthenticated firmware.

Accelerator designs are generally proprietary and Secure Boot requires firmware changes to be signed, so firmware updates would likely require substantial assistance from chip designers. Several different manufacturers make components that are on each accelerator, so this would not necessarily require assistance from the primary accelerator designer.

## Software Modifications

The most straightforward strategy for third-parties to execute guarantee logic is to create software that runs on the CPUs that are responsible for managing the accelerators. Verification of the software that is run can be provided by Trusted Execution Environments (TEEs) [15], which use a hardware root of trust to provide difficult-to-fake evidence that the computing environment is configured as claimed. Mithril Security [16] and EQTY Lab [17] are currently working on strategies like this with the goal of providing a credible record of how a model was trained. EQTY's whitepaper describes their approach in more detail, but at a high level, hardware security keys from the CPU and GPU are used to sign a record of the software initially loaded on the device, and a session key that signs any data or code that is later loaded for training. If an auditor is provided with detailed logs containing all of the software and data used, these signatures provide evidence that the software was executed as claimed in a specific computing environment. The signatures are difficult to fake because the hardware keys are stored in secure hardware memory.

At the time of writing, Nvidia has released an early-access implementation of Confidential Computing (based on a TEE managed by the host CPU). The early access implementation supports a single node with AMD or Intel CPUs, but does not yet support multi-node or Grace CPUs (which limits the possibility for immediate widespread deployment). Nvidia's Confidential Computing implementation doesn't currently claim to be secure against sophisticated physical tampering. Additionally, security is one of many sometimes conflicting priorities for high-performance CPUs, and significant non-invasive TEE vulnerabilities are sometimes discovered [19], [20].

---

[5] There are also common defenses against fault injection attacks, like repeating important steps several times in series or in parallel to be able to detect inconsistencies.

Using a TEE makes it more difficult to falsify workload records, but does not prevent developers from using different software without any guarantees built in. For this reason, software-based solutions could enable verification about how hardware has been used, but would not be able to enforce that the hardware cannot be used in guarantee-breaking ways.

TEE-based proof of training could be implemented in the near-term by sharing workload logs with auditors so they can manually check that the workload is compliant. More advanced designs may be able to automate this verification process so that AI model details do not have to be revealed. These automated checks could be done in a few ways, including: 1) providing developers with an API for model development that blocks non-compliant workloads 2) inferring workload characteristics by scanning code and gathering usage data 3) allowing developers to make specific claims about why their workload is compliant, and more narrowly verifying these claims. We explore these options in more detail in General-Purpose Guarantees. If the workload passes the check, the session key can be used to sign a certificate stating that the model passed without needing to reveal additional details. Progress on automated guarantee checks in software would likely translate to hardware-based solutions if the same workload information is available. With a lower barrier to entry, this makes TEE-based software a promising development environment for automated guarantee checks that are eventually intended to run on dedicated hardware.

Overall, TEE-backed software for workload verification is relatively fast to deploy, but probably not secure against sophisticated actors if they have unrestricted physical access. This limited security may be sufficient for domestic regulation, where it could be difficult to hide major efforts to attack verification mechanisms. In international contexts, TEE-backed software could provide verification capabilities while more trusted hardware is developed. This could be made more secure if combined with tamper evidence, national intelligence agencies searching for circumvention attempts, and additional sensors collecting measurements to corroborate claims.

## Accelerator Measurements

Several data sources could be used to verify workload guarantees, with different sources enabling different types of guarantees. The following table lists potential data sources and their respective advantages.

| | |
|---|---|
| **Accelerator Memory (HBM)** | Direct memory access would provide visibility into workload state. This access could enable verification of claims about model architecture and facilitate snapshots of accelerator state for auditing. |

| Accelerator Instructions | If the accelerator is trusted to execute instructions, monitoring the kernels sent to the accelerator could enable analysis of workload characteristics. One limitation |
|---|---|
| Network Traffic | For distributed workloads, monitoring inter-accelerator communication could provide important information about the overall computation being performed. This could also be useful for guarantees about the network configuration. |
| Performance Counters | Hardware performance counters could provide metrics such as FLOP counts, memory access patterns, and interconnect utilization. |
| Power Usage | Power consumption monitoring offers a data source that is difficult to falsify without physical modifications. Power signatures could help identify workload types and detect anomalous computation patterns. Additionally, power measurements could serve as a cross-validation mechanism to verify consistency with other reported metrics. An existing FlexHEG prototype uses high resolution power measurements. |

Data sources that are consistent across different accelerator types are preferable, as they reduce the amount of customization needed for each hardware platform. Using multiple independent data sources could improve the robustness of guarantees, since attackers would need to falsify all sources to avoid detection. The availability of these measurements depends on how the Guarantee Processor connects to the accelerator, while their trustworthiness is determined by whether the measurement occurs within the Secure Enclosure and how much it relies on components that are vulnerable to attack or difficult to audit. As discussed in the [Interlock](#) section, placing the Guarantee Processor on a critical data path for accelerator usage is one option to improve the trustworthiness of measurements.

# FlexHEG Hardware Components

In this report, we mainly focus on flexHEG implementations that use a Guarantee Processor to collect accelerator data and perform guarantee checks, and a Secure Enclosure to protect the guarantees from being circumvented with physical tampering. The Guarantee Processor and Secure Enclosure are intended to function separately from each other to make the design and deployment process simpler, and also because in some scenarios it may be important for separate parties to design and manufacture these components. This section analyses requirements and tradeoffs for the Guarantee Processor and Secure Enclosure.

# Guarantee Processor

With an on-device Guarantee Processor, checks can be done locally in a privacy-preserving way where sensitive data does not need to leave the device. The Guarantee Processor could simply provide a cryptographically signed certificate saying that the guarantee has been met, which the device owner could use as credible proof if they choose to. Additionally, a dedicated guarantee processor can be much more auditable than a full accelerator system, because it can be smaller, simpler, less performance sensitive, and would be possible to open source without revealing trade secrets. This is especially advantageous for being able to add it to untrusted accelerators.

As previously discussed, it is important that the Guarantee Processor have robust access to accelerator measurements. In addition, the guarantee processor ideally:
- Is performant enough to execute the guarantee calculations discussed in [General-purpose Guarantees](#). For example, it would be useful for the Guarantee Processor to be able to double check the work of one Streaming Multiprocessor (SM) on a Blackwell chip, which would require roughly 1/192 the computational power of a Blackwell chip (because they have 192 SMs)
- Is secure against cyber attacks that might attempt to make use of flaws in the implementation.
- Is secure against physical attacks that might make it through the secure enclosure
- Is able to perform necessary cryptographic algorithms, and can do so without revealing information via side channels like timing or power. A special case of this is when the Guarantee Processor is responsible for efficiently encrypting the data stream, as discussed for the [Interlock Cryptography Hardware](#). Public key cryptography is also needed for verifying the authenticity of updates and for signing certificates about which guarantees have been met. This public key cryptography is ideally quantum secure (e.g zeroRISC's work for Secure Boot [22])
- Is auditable by third parties that might not trust the designers or manufacturers
- Is possible to update if and only if authorized by the required parties (see [Guarantee Update Process](#)).
- Has access to a reliable real-time clock for time-dependent guarantees.
- Has access to a secure random number generator.

There are several existing processor IP blocks that could be used to satisfy at least a minimal version of these requirements, including a range of TPM cryptoprocessors [21]. As an example of a related (although closed-source and obfuscated) processor, the Intel Management Engine is included on many Intel CPUs for features like anti-theft prevention and capability licensing [22]. There are also open-source alternatives [23], [24], [25].

To increase trust that backdoors have not been added to the Guarantee Processor design or manufacturing process, some options are:

- Open-sourcing the guarantee processor design and guarantee logic, which could then be analyzed and compared with physical scans of randomly selected chips.
- Using FPGAs to execute some or all of the guarantee logic, because they can have their configurations publicly audited, might be easier to check for physical inconsistencies because of their more uniform structure, and can use variants of configurations with equivalent logic to make many hardware trojans impractical [26].
- Sourcing multiple guarantee processors from different manufacturers[6] and running the same instructions on all of them before checking their outputs for consistency on-device (this approach is similar to the Lock-Step, which is used to provide fault tolerance in functional safety designs [27]). With three guarantee processors, they could detect disagreements and use ⅔ voting to choose which action to take.

The guarantee processor should be able to defend against voltage glitching and similar attacks [28]. There are also manufacturing techniques that would probably make tampering harder if an attacker is able to circumvent the secure enclosure (e.g., putting the guarantee processor near sensitive or fragile components like tamper sensors or interconnect channels). The FIPS-140-3 certification measures the security of similar hardware security modules, and a similar certification process could potentially be used for the Guarantee Processor.

A dedicated Guarantee Processor could be added to future accelerators as a standalone chiplet, or as a part of another die. Alternatively, for a retrofittable solution, it could be added to existing accelerators as a plug-in device (e.g. using a PCIe slot that can be configured to read from HBM with Direct Memory Access (DMA) [9]).

As mentioned in the Firmware Modifications section, several components on existing accelerators already have general-purpose processors that could be updated with manufacturer assistance. Repurposing existing components would make deployment much easier, but may require compromising on some of the previous requirements. Some options for components to repurpose are analyzed in the table below.

| Processor(s) on AI chip | **AI chips** often have reprogrammable processors directly on the main die. The designs are generally proprietary, but these processors are likely used to handle tasks related to general chip management, security and data processing. Nvidia reportedly has 10-40 cores on each GPU that are based on the open-source RISC-V architecture [29]. Some of these RISC-V cores power the GPU System Processor, which is responsible for communication with the CPU kernel driver, and "has full access to everything in the GPU, including access to the memory controllers". Security logic on these microcontrollers appears to be formally verified [30]. |
|---|---|

---

[6] Although this may increase the risk of a single compromised guarantee processor being used to sabotage chip usage.

|  | |
|---|---|
|  | Processors on AI chips are especially promising for implementing guarantees because they have direct access to the chip state, and being on the same die makes them more difficult to tamper with. |
| **High bandwidth memory** | **High bandwidth memory** (HBM) is used in AI accelerators to store workload state (e.g. model weights and gradients). Currently HBM doesn't have significant (or potentially any) on-chip reprogrammable computing ability, but major HBM designers (e.g. Samsung and HK-Synx) are attempting to add it. HBM-PIM (Processor In Memory) and HBM-PNM (Processor Near Memory) are two proposed designs that would allow some accelerator computations to be performed without leaving the HBM block, which could improve efficiency.<br><br>If on-die processors were added to future HBM chiplets, they would be a promising option for implementing flexHEG logic because:<br><br>● They have direct access to workload state<br>● HBM is difficult to manufacture and is an important part of all leading accelerators<br>● They would be simpler and more transferable between accelerator types than many other components in the accelerator system<br>● They could take snapshots of memory<br>● They could randomly check accuracy of a fraction of computations and data transmissions to verify that other untrusted hardware is acting as claimed |
| **Network interface controller** | **Network interface controllers** are used to handle communication between each accelerator and the scale-out network. Modern NICs like Nvidia's ConnectX-8 have programmable compute capabilities for "in-network" processing and hardware for inline cryptography, while being able to process up to 800Gb/s (100GB/s) of data. NICs are an important part of datacenter security because they can encrypt data that leaves each node, protecting data in transit from being read or modified.<br><br>NICs can directly observe the data sent through them to the broader network. They can also use Direct Memory Access (DMA) to read the contents of HBM.<br><br>The NIC's architecture is typically simpler than a CPU, which can reduce its attack surface, and its position at the interface between the network and compute node creates a natural security boundary. Amazon's Nitro system leverages this approach [31], using NICs as hardware-backed security boundaries to isolate virtual machines by offloading security-sensitive |
| | operations. Other companies implement similar security strategies with NICs, though with less public documentation.<br><br>NICs are well situated to observe workloads that involve many accelerators because they sit on the main data path for communication between accelerators across nodes. This strategic position, combined with the ability to swap out the NIC with one made by a different manufacturer (as Amazon is doing with their NVL72 systems [32]), makes NICs a strong candidate for implementing flexHEG logic. |
| **Datacenter Secure Control Module** | Datacenter Secure Control Module (DC-SCM) is a standard for a module used to perform server management and provide security functionality [33], [34]. NVL72 is mostly compatible with the DC-SCM standard (the form factor is slightly smaller but the electrical interfaces are the same[7]).<br><br>DC-SCM is plugged in via PCIe rather than built directly into the motherboard like typical baseboard management controllers (BMCs). This allows reuse of components across different motherboards, and enables data center operators to replace the module without replacing the entire motherboard (e.g., if a hardware update is needed for security reasons). DC-SCM might be a relatively simpler option for implementing flexHEG logic because:<br>● It is less critical for accelerator performance, so might be more feasible to modify without disruption<br>● It would be relatively easy to replace with a dedicated flexHEG version<br>● They already have access to information about the board like power usage<br>● As a PCIe device, it may be possible to configure read access of HBM<br>One downside of DC-SCM modules, however, is that they are not on the main data path, so data readings may be easier for an attacker to spoof. |

Other components on or connected to accelerators within data centers could potentially be repurposed for flexHEG, including network components like switches. It is also an option to repurpose multiple components, which could provide defense in depth and more comprehensive information (although for simplicity, we will not discuss this option more here). To make auditing and design simpler and faster, it may make sense for a submodule within a component to be used for flexHEG, rather than the entire component.

---

[7] "We also developed a new, denser DC-SCM (Data Center Secure Control Module) design that's 10% smaller than the current standard" [35].

In the Interlock section, we analyze a specific design where data entering and leaving the accelerator must pass through the Guarantee Processor (or rather a buffer that it has direct access to).

## Secure Enclosure

A secure enclosure is used to protect flexHEG devices from physical tampering. This is needed, because otherwise an adversary could modify hardware to circumvent guarantees, or use measurements to exfiltrate confidential data. Depending on the threat model (see [Appendix B](#) for more on threat models), there are different ways to dissuade tampering, including using a tamper evident enclosure, or a tamper-responsive enclosure that triggers disablement of the accelerator. Another option that is a mixture of tamper evident and tamper responsive is to incorporate a Physically Unclonable Function (PUF) in the enclosure, where tampering with the enclosure would corrupt the measurements used to derive a secret PUF key. Without the PUF key, adversaries would not be able to sign attestation certificates or decrypt important data. The secure enclosure boundary could potentially contain a single chip, a single accelerator PCB, several accelerators within a tray, many accelerators within a rack, or even larger configurations. The choice of where to put the security boundary has several tradeoffs between security, maintenance, and time-to-readiness. It also influences where flexHEG logic can be securely executed, and where encryption is needed (e.g. if chip-level security boundary, ideally want to encrypt data leaving the chip). In this section, we summarize publicly available information on hardware security that is relevant for AI accelerators and flexHEG threat model, and also how trade offs here influence the overall flexHEG design.

Secure enclosures have been used for decades to defend cryptographic coprocessors and other chips from physical tampering [36], [37]. The FIPS-140 certification process [38][12] measures the physical defenses of tamper-responsive enclosures for cryptographic coprocessors. At the time of this writing, there are four devices with the maximum FIPS-140 rating (level 4) that are sold by IBM [14] and by Private Machines [15]; neither of the systems have much public information about their tamper resistant security features, likely because security-by-obscurity is commonly practiced in hardware security, which is more offense-dominant than cryptographic security. The IBM and Private Machines devices have not been publicly broken[8], although it is unclear how secure they would be against state-backed actors. There are also rumors of private US government projects on hardware security that may have even more advanced defensive measures. Hardware security from these related domains could potentially be translated to work with AI accelerators if it can be modified to accommodate the accelerator form factor(s) and to handle high heat dissipation.

Tamper-evident enclosures could be useful in scenarios where accelerators can be inspected and where misuse does not need to be prevented immediately. Building tamper-evident enclosures could potentially be faster and simpler than tamper-responsive enclosures (they probably do not require a battery or self-destruct mechanism). Techniques used for monitoring nuclear materials, like tamper evident seals and 24-7 video cameras could be fairly simple to adapt to this use case (and data centers typically already have these cameras). Some of the older tamper-evidence techniques may not be perfectly secure, like the cold-war seals that can apparently be broken [40]. However, this type of tampering would probably be difficult to perform on a large fraction of accelerators in a data center without being detected. The security could also be improved by layering different types of tamper resistance and tamper evident defenses.

---

[8] Though an older IBM cryptographic coprocessor was broken [39].

Physically unclonable functions (PUFs) are usually used to provide persistent secret information that cannot be passively measured with scanning equipment. PUFs do this by taking measurements of unique hardware characteristics and using them to generate a secret key. This is normally done with components like SRAM PUFs [41], but the same principle could be used for a secure enclosure. For example, the PUF could be based on electrical properties of a conductive mesh that is embedded in the enclosure wall.

An advantage PUFs have over active sensing is that they don't require a battery. However there are also a few limitations to PUFs:
- The PUF key generation procedure might take seconds, so might be too slow to respond to some attacks without also pairing active sensing (closer to millisecond response time)
- To use PUFs to prevent future usage of accelerators that have been tampered with, the destruction of the secret key must prevent decrypting data that the accelerator cannot operate without. This could potentially be the firmware or tuned coefficients for the accelerator, which would make it difficult for adversaries to use the accelerator if they do not have a functioning version of the firmware or tuning procedure. Against sophisticated adversaries, this may not be sufficient though, because they would only need to acquire this firmware once to unlock all of their devices, and this firmware could potentially be leaked by a single rogue employee that works for the accelerator manufacturer.
- While PUFs make passive scanning of secret information difficult, when that secret information is read in order to recreate the secret key, that information is transmitted on data buses and stored in registers. These registers and buses could potentially be read with microprobing or laser voltage probing. One strategy used to mitigate this is using a "strong PUF" that has enough entropy to generate many keys, so that the same information doesn't need to be read repeatedly. However, being able to generate many different keys does not immediately solve the problem, because the secret information on device needs to be encrypted and decrypted with a particular key, so to do this locally would require multiple reads of the same key. A different defense that enclosure PUFs could use against active scanning of registers is to design the physical enclosure in such a way that signal wires and registers are not possible to access for scanning without first destroying the PUF.

Tamper responsive enclosures use active sensing to detect if the enclosure has been breached and then to trigger an appropriate response. One approach for active sensing is to measure the capacitance of a serpentine-patterned conductor embedded around the enclosure [9]. Similarly, the resistance of a conductor mesh can also be measured [10]. If the mesh is disrupted, then the measurements will change, alerting the tamper-detection system. Measurements of the radio response function within an enclosure could also be used [11] (although detection tests for the referenced prototype have so far only been done with metallic probes). Additional sensors can be used to prevent more sophisticated attacks by measuring for radiation, voltage glitching, lasers, temperature, and the rate of temperature change. To avoid having a single point of failure that could be targeted by an attacker, a distributed network of sensors could be used to independently trigger the tamper-response mechanism.

The appropriate tamper response mechanism depends on the threat model. If the goal is to prevent confidential data from being stolen, it is important to delete all data and secret keys when tampering is detected. Additionally, it is also important to communicate that this tampering has happened, so that the same accelerator is not used again later after it may have been compromised. One option for this is for each accelerator to frequently send an "all OK" signal, and to investigate accelerators that do not send this signal on schedule.

If the goal is to prevent tampering being used to circumvent guarantees about future accelerator usage, tampering should trigger permanent disablement of the accelerator before the guarantees can be disabled. There are multiple approaches that could be used to disable an accelerator. Some options are:

- **Capacitors for voltage-based destruction**: Large capacitors could be charged during normal operation and then discharged to apply excessive voltage to critical components of the chip when tampering is detected. This voltage surge would exceed the chip's electrical tolerance, causing permanent damage to vulnerable circuit elements. This approach could be implemented with relatively simple circuit additions to existing designs and doesn't require changes to the main accelerator die, making it a good option for retrofitting.
- **Embedded antifuse bits**: Modern chips can include one-time programmable (OTP) memory cells based on antifuse technology. Unlike regular fuses that start connected and can be blown to disconnect, antifuses start disconnected and can be permanently connected by applying a high voltage. By embedding thousands or millions of these bits throughout the accelerator chip and designing the chip to require specific antifuse configurations to function properly, tampering detection could trigger widespread antifuse activation that would render the chip permanently inoperable. This approach is highly effective but requires deep integration during chip design and manufacturing, making it unsuitable for retrofitting to existing accelerators.

- **Nanothermite-based destruction:** A more experimental approach involves embedding reactive materials like nanothermite (a material that can produce a rapid exothermic reaction) on the chip surface. If tampering is detected, the nanothermite could be ignited by a small electrical pulse, generating intense localized heat that would physically destroy critical chip components.

To ensure system integrity over the lifetime of the device, a battery inside or outside of the enclosure can be used so that the sensors continue to operate even if external power is disconnected.

Cooling is necessary for AI accelerators, which have to dissipate up to 1200W of heat per chip (Nvidia B200). Most existing data centres use fans to air cool chips, but some next generation accelerator racks like the NVL72 require direct-to-chip liquid cooling in order to achieve the intended compute density (trays with air cooling are taller than trays with direct to chip liquid cooling). High compute density allows for shorter interconnect cables, which reduces communication latency and cabling cost. Secure enclosures can ideally be made compatible with both air and direct to chip liquid cooling (and potentially two phase direct to chip cooling which is speculated to be the next advancement [42]).

To permit cooling, the secure enclosure boundary can either be sandwiched between the chip face and cooling pad, or it can contain both the chip and the cooling pad. If it is sandwiched beneath the cooling pad, it has to be very thin and very thermally conductive. Alternatively, if it is further away from the chip face, the secure enclosure has to allow heat to flow through it. For air cooled units, this would probably have to allow airflow to pass through the boundary. The security risk of having holes in the enclosure could potentially be reduced by making the air pass through a metalized foam that doesn't permit a direct line of sight into the enclosure, and whose particular geometry is sensed by radio. For direct to chip liquid cooling one option is to have the liquid tubing pass through the enclosure boundary, potentially with a similar defensive layer that allows liquid though. Alternatively, the liquid tubing within the enclosure could form a closed loop, and heat could be exchanged at the enclosure boundary with a liquid-liquid heat exchanger.

Larger enclosures that contain more accelerators are likely faster to design and deploy because they do not need to be as tightly integrated with the specific accelerators (as an extreme example, the "secure enclosure" could be an entire datacenter). Ease of deployment is further improved if the secure enclosure matches standard dimensions for a server tray or rack, like the open sourced MGX rack standard used by NVL72 [43]. However, enclosing multiple accelerators has potential downsides:
- The more accelerators in each secure enclosure, the higher the probability that one of them will fail or require maintenance, and for security reasons it would be preferable to not be able to open the enclosure after it has been closed

- An attacker would only need to disable one enclosure to access multiple GPUs, and a larger surface area has more room for flaws

To put many accelerators within a single secure enclosure that cannot be opened would be feasible if hardware failure requiring maintenance or replacement was extremely rare. Alternatively, the number of accelerators within the secure enclosure could be over-provisioned, so that the combined system could still function after a few hardware failures (as is done with solid state drive memory regions). The feasibility of over-provisioning depends on the software and network architecture. This is simpler for hub-and-spoke models like NVL72 than for network grid layouts like Google's TPUs, where a single failed device would cause non-trivial issues with network routing.

As previously mentioned, the security of the guarantee processor and the measurements it makes is dependent on the secure enclosure boundary. A Guarantee Processor on the accelerator chip that makes chip-level measurements could potentially be protected with a chip-scale secure enclosure. This has the advantage that it can be included early in the supply chain. It would also likely require advanced tooling (like a focused ion beam machine) to physically tamper with logic on the chip. A defense in depth approach is probably useful here, because the addition of a larger secure enclosure could make it more difficult for attackers to get tooling into place to tamper with the chip enclosure.

To summarize, there are many options for adding secure enclosures to accelerators, with different tradeoffs. Based on the public literature, it is very difficult to circumvent the best secure enclosures. However, nation-state attackers can likely compromise the best current secure enclosures, but it is difficult to estimate what the marginal per-device cost of circumvention would be.

# Interlock-Based FlexHEG Design

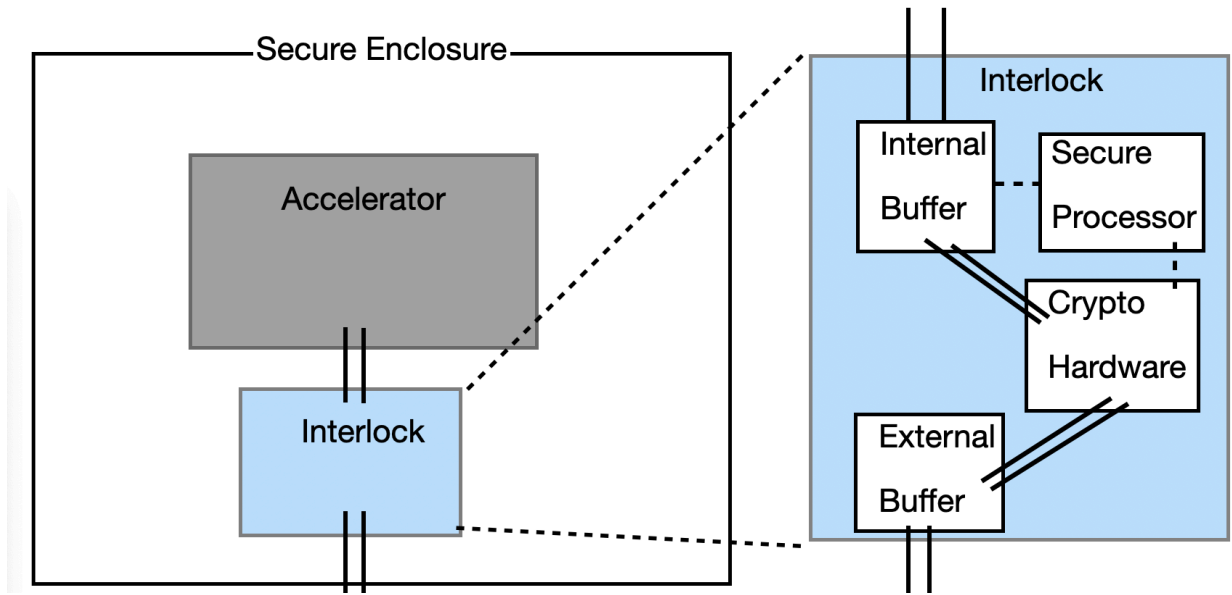

Figure 4: Interlock-based flexHEG design

As previously discussed, integrating flexHEG with core accelerator functionality would make it harder for attackers to spoof data or circumvent guarantees. One promising approach is to position a component on the main data path. We refer to a flexHEG component sitting on the data path as an "Interlock."

A high-level visualization of this design is shown in Figure 4. While the figure shows a single accelerator with a single interlock in a secure enclosure, multiple accelerators could alternatively be held within each secure enclosure. As the only path for data and instructions to enter the accelerator, the interlock would be able to robustly monitor and/or control this flow.

The interlock can use encryption to protect the data that leaves the secure enclosure from being read or modified by an attacker. By controlling the key exchange process used to initialize encrypted channels between nodes, the Guarantee Processor would have clear visibility into the cluster configuration.

Understanding data flows between accelerators is very important for making high-level guarantees about distributed workflows (as each accelerator in isolation may appear to be doing standard matrix multiplications). Using an Interlock is a relatively straightforward way for the guarantee processor to see how data is moving between accelerators. Although, if needed, there are other options to check the authenticity of incoming data, such as comparing the logs of data in memory between accelerators.

A downside of having the Interlock on the main data path is that performance requirements are higher, and it may be required to upgrade to meet future performance requirements (compared to a passive measurement device that doesn't need to be modified as frequently). The ideal implementation would have direct access to a simple part of the data path, like a buffer, and direct control over crypto hardware, but not have to implement all of the logic and performance optimization. An easier to implement/scale version would be for the interlock to not perform the inline encryption, and just be able to read the buffer.

A benefit of the interlock design is that it can treat the accelerator as a black box, with less need to understand or trust internals, which could speed up adaptation to novel accelerators and require less information sharing from accelerator designers. Since the Interlock already sits on a data pathway, it could likely also separately use DMA to read internal accelerator data.

The Interlock could be implemented in several ways, including 1) as an IP block within the accelerator die, 2) as a chiplet sitting near the accelerator die, 3) as a part of high bandwidth memory (HBM), or 4) as a separate component that operates as a network relay or switch. The space constraints for putting the Interlock on-die or near the accelerator chip and HBM are more demanding because this physical layout affects memory latency.

Relatedly, the requirements for Interlock latency and throughput are more demanding the closer the interlock is on the data path to the accelerator die, because accelerators typically access HBM much more frequently than the cluster network (e.g. H100s expect ~3TB/s of HBM data transfer, compared to 900GB/s NVLink, or 128GB/s PCIe to route to the broader network).

## Repurposing the Network Interface Controller

A promising option for a retrofittable Interlock is repurposing or replacing the Network Interface Controller (NIC), which as previously discussed is responsible for communicating with the scale-out network. FlexHEG designs that are external to accelerators are easier to physically retrofit to novel accelerator types, or integrate later in an accelerator design cycle.

This approach draws inspiration from existing security architectures like Amazon's Nitro system [31], which uses the NIC as a hardware-backed security boundary to isolate virtual machines by offloading security-sensitive operations to their Nitro Card.

A potential challenge with updating the NIC is that hyperscalers already have protocols for configuring clusters within their data centers, and custom hardware for network management and encryption. It would likely be possible but costly for them to replace these technologies with a flexHEG version.

# Encrypted Cluster Formation

FlexHEG systems need to be able to securely communicate with each other so that they can coordinate around guarantees, and so that data in transit can't be read by adversaries. To enable this, flexHEG systems could follow the encrypted cluster formation protocol described in the table below. This protocol combines public key cryptography for initial authorization with efficient symmetric encryption for ongoing data transfer, allowing clusters to be formed securely while maintaining the high-bandwidth, low-latency communication needed for distributed AI workloads.

1. The (directed) network graph of possible communication pairs is initially declared (with unique device IDs for each endpoint).
2. The guarantee processor at each node parses this graph to assemble a list of the IDs of all nodes that it needs to send data to.
3. Two options for endpoint key lookup:
    a. Logging/approval version: Each guarantee processor sends a request to an external governance system, asking for permission to send data to the list of endpoints, and optionally for any bandwidth or time limits. The governance system sends an approval certificate and the public keys of the requested endpoints.
    b. Decentralized version: each guarantee processor checks a local lookup table[9] for the public keys of the requested endpoints.
4. Each guarantee processor randomly generates a session key for AES-GCM encryption and loads it into its output-path crypto engine.
5. Each guarantee processor encrypts its session key with the public keys of approved endpoints, then sends this data to those endpoints.
6. All guarantee processors decrypt the session keys of source nodes and optionally check for approval to communicate with these nodes (this could be the same certificate from step 3a).
7. The approved session keys are loaded into input-path crypto engines to decrypt incoming data.

Besides protecting communication channels, this protocol could optionally provide mechanisms for governance of cluster configurations. If option 3a is chosen instead of 3b, all flexHEG devices could be required to report cluster configurations before starting a workload, which could be a useful level of transparency for international coordination. Routing network approval through a more centralized service would also reduce the amount of public-key cryptography that needs to be done, which may become significant if direct all-to-all communication is needed in very large clusters.

---

[9] This lookup table would typically be provided by the accelerator manufacturer (Nvidia supports this now for air gapped checks of device attestation [44]). For multilateral use cases it would be helpful to have a multilaterally maintained service. The easiest way to do this would probably be to have each party that monitors the flexHEG manufacturing process publicly share their own table. Third parties could check all tables for consistency.

To allow network switches to run computations on transmitted data (e.g. SHARP [45]), network switches would have to also be set up as flexHEG nodes. They could then decrypt the data, run any computations, then encrypt and transmit the resulting data.

AI workloads frequently involve one-to-many communication, which is why this protocol uses a single symmetric key for each source node (instead of a unique symmetric key for each node pair). A limitation of this is that if one flexHEG node is compromised, the source keys it has could be used to read or spoof transmissions from those source nodes. One way to mitigate the risk of spoofing would be to occasionally sign the hash of previously transmitted hashes using the private key rather than the symmetric key (although care needs to be taken to handle dropped packets sensibly).

## Cryptography Hardware

Cryptography that provides confidentiality could allow the guarantee processor to choose specific accelerators that are allowed to read the outgoing data. Similarly, cryptography that verifies message authenticity could allow guarantee processors to trust that a particular message has come from another guarantee processor.

There are several potential applications of encrypted interconnect, including:
- Protect weights from being exfiltrated by an attacker snooping on interconnect;
- Constrain a training run to only use a certain, limited number of GPUs (similar to Fixed Set [11]) by only allowing specific GPUs to decrypt the information;
- Verify authenticity of metadata describing the compute graph, which could enable verification that training data were approved by a regulator or that model parameters have not exceeded an operation limit.

A common algorithm used to provide Authenticated Encryption with Associated Data [46] is AES-GCM [47]. AES-GCM operates in counter mode, which means encryption of a single data stream can be efficiently parallelized because each block is encrypted separately. A security risk with AES-GCM is that the nonce cannot be re-used even once, as this could allow attackers to infer the secret key [48]. The cipher for AES-GCM can be precomputed prior to seeing the data, and only needs to be XOR'd with the data stream to compute the encrypted output stream. This allows specialized hardware to perform the AES portion of the encryption with close to 0 latency, and then perform the follow-up GCM calculation for authenticity and send this result separately [49].

> Order-of-magnitude estimate for fraction of compute needed for 256 bit AES-GCM encryption on an H100:
> - H100 capable of 1.98 *10^15 int8 ops per second;
> - Interconnect bandwidth of 900*10^9 bytes per second;
> - ~ 50 operations per byte for AES256;
> - AES-256 applies roughly 14 rounds with 4 operations each to all input data, so roughly 50 operations per byte. Authentication with GCM typically takes half the time as the AES component , so we will approximate as roughly 75 operations per byte in total (although exact performance depends a lot on implementation details, and could maybe be accelerated with dedicated hardware)
> - So to encrypt all NVLink traffic would require 900*10^9 * 75/(1.98 *10^15) = ~ 3% of computing power / energy / board space;
> - To encrypt all HBM would require 3000*10^9 * 75/(1.98 *10^15) = ~11% of power;
> - Or to encrypt all PCIe traffic would be 3000*10^9 * 75/(1.98 *10^15) = ~0.5%;

To perform the cryptography for authentication and confidentiality, the interlock ideally has a hardware-accelerated crypto engine that supports high bandwidth with minimal latency. An additional benefit of cryptography hardware is that it can be specifically designed to minimize information leakage via side channels that may be exposed by software implementations. Such cryptography engines are common in devices like solid state drives or network cards that need to perform inline encryption (although typically they don't need to support as much bandwidth as is being considered here). Alternatively, this cryptography could potentially be performed by modifying GPU kernels to encrypt data after reading and before writing to HBM, however safe transport of keys would need to be solved.

The NVIDIA H100 can provide encryption of PCIe traffic to support single-GPU confidential computing, although this specific implementation imposes additional latency [50]. Nvidia has suggested that NVLink may be possible to encrypt in Blackwell GPUs, although no public details are available on this at the time of writing.

Inline cryptography hardware is common in modern Network Interface Cards [51], which offer up to 800Gb/s and typically connect accelerators to the frontend or backend network [52]. Modern NICs also typically have general-purpose processors, which are discussed more in the section, A Specific Retrofittable flexHEG Design.

If the interlock does not have enough computational power to encrypt all traffic that passes through it, partial authenticity could be provided at a small fraction of the cost using pseudo-randomized authenticity checks, as described in [Appendix C](#).

# General-Purpose Workload Guarantees

This section discusses approaches to implementing general-purpose verification of guarantees, ranging from simple logging to automated verification of complex guarantees. The feasibility of implementing and securing these approaches for guarantee checking depends on the flexHEG hardware, as discussed previously. The strategies discussed here are presented at a high level because some details depend on the specific type of guarantee. In [FLOP Counting](#), we discuss a more specific design for multi-accelerator guarantees about total FLOPs.

We analyze four potential approaches to verifying claims about computations:
1. Transparent logging and auditing: The system maintains detailed logs of all operations, which can be manually audited. This approach is suitable when sharing detailed information about computations is acceptable.
2. Automated scanning of unmodified workloads: The system attempts to automatically classify workloads as compliant or non-compliant by analyzing code, usage patterns, and other observable properties.
3. Compute graph declaration: Developers make specific claims about their computations, either through upfront static declarations or dynamic operation-by-operation declarations, which the system can verify. The system compiles these declarations into a verified compute graph, which can be more robustly parsed for particular workload guarantees.
4. Framework-based verification: The hardware is only made accessible through a framework with a limited set of APIs. These APIs either structurally constrain workloads to follow guarantees, or make it much simpler to automatically check guarantees (e.g. by checking parameters that have to be entered into the API).

## Transparent Logging and Auditing

For use cases where detailed auditing is feasible and acceptable from a privacy perspective, approaches based on logging and snapshots can be effective, as described by Shavit [53]. To accomplish this, memory snapshots could be taken at random points in time, with the hashes of these snapshots stored securely in hardware. Then during auditing, the developer would provide full memory snapshots that match the hashes, along with the code and input data that explain the progression between snapshots. Optionally, other information about the system could also be logged to assist auditing, like the kernels executed or network communication patterns.

While relatively straightforward to implement (compared to fully automated guarantee checks), this approach requires sharing potentially sensitive information about computations. Manual audits might take a lot of time from skilled engineers to ensure that workloads were not subtly designed to hide non-compliance, especially if the architectures used are substantially different between developers. This time requirement could potentially be reduced with tooling to assist the manual process.

Alternatively, the logs could be stored locally and not exported unless they need to be checked after an incident (somewhat like a "black box" on airplanes). Or as a potential in-between, regulators could get access to logs after a set delay so that developers are not as worried about leaking cutting-edge secrets.

## Automated Local Scanning of Unmodified Workloads

For use cases requiring privacy-preserving verification, it is likely possible to automatically verify certain properties of workloads without requiring any modifications from the developer. This approach is most feasible for guarantees about properties that can be directly measured or inferred from observable system behavior, such as:
- Total clock cycles used
- Power consumption
- Number of accelerators involved
- Basic communication patterns between accelerators
- Memory usage patterns

Developers could additionally be required to provide the Guarantee Processor with the accelerator kernels or other detailed information about the workload. Even if it is not possible to directly verify the authenticity of some data, it may be challenging for an adversary to keep complex claims consistent across multiple data sources. It may be possible to use this combination of data sources to predict more abstract characteristics about the overall workload (e.g. is it doing reinforcement learning).

Since this type of check is based on general patterns associated with certain workloads and not exact logic, it could suffer from false positives or false negatives depending on how the classification is tuned. If there is a false positive, a developer may not want to reveal their true workload to prove compliance. A potential solution might be for the developer to provide a minimal example of a false positive on a different compliant workload. To find false negatives in the classification algorithm, overt adversaries could iteratively modify their noncompliant workloads until they are classified as compliant. In some situations, the reduced efficiency these modifications impose on non-compliant workloads may be worthwhile, even if the check is not fully robust.

If an adversary does not want to be detected breaking any guarantees, and does not know the details of the workload check, it could be much more challenging for them to falsely mimic a compliant workload without risking detection. This difficulty depends on the classification logic being kept secret, which would be challenging to do against state adversaries, and would not be feasible for some types of agreements that rely on openness of the mechanism. One strategy that may dissuade covert adversaries would be to periodically update the classifier, and then parse older usage logs with the updated classifier to see if any past breaches had been missed.

Another option for performing automated checks is for the check to itself be done by a large language model (LLM) that is run locally. Running the LLM locally would provide credible evidence to the developer that additional information was not being exfiltrated - this could either be done on the same hardware that is being monitored, or on additional secure hardware dedicated for this purpose. This LLM could review kernel logs and usage statistics, and check for more sophisticated attempts to camouflage non-compliant workloads. As AI capabilities improve, the LLM could be updated to more intelligently parse past and future logs (again, covert adversaries would have to risk that future updates would not discover any previous deception).

Empirical work is needed to create a dataset of compliant/noncompliant workloads and develop classification algorithms. Red teaming and usability testing of candidate classifiers would provide more information about real-world usefulness.

## Compute Graph Declaration

This approach requires developers to make detailed claims about the operations that their workloads are composed of, which the flexHEG system then automatically verifies for compliance with high-level guarantees. These claims would be composed of atomic operations acting on chunks of data, which would include basic mathematical operations (e.g. matrix multiplication), and data transfer (potentially between nodes). The set of ordered operations that are performed can be constructed as a compute graph[10] that describes all steps in the chain of events that create each chunk of data.

These claims can be made in two ways:
1. Static declaration: The developer provides upfront a complete description of the intended computation, including all data transfers and computational operations.
2. Dynamic declaration: The developer declares each atomic operation (either a data transfer or computation) as it is performed. The system then verifies that each atomic claim is actually executed as expected and logs the series of operations that occur.

With static declarations, the flexHEG system could analyze the entire workload for compliance before execution begins, and then verify that what is executed matches the claimed workload. With dynamic declarations, the flexHEG system could periodically check during execution that the recently logged compute graph is in compliance with workload level claims.

---

[10] See pytorch and Nvidia docs for examples of current compute graph implementations in frontier AI workloads [54], [55], [56].

Static declarations would be suitable for workloads that don't change the control flow based on data, like performing a fixed number of inference steps with a dense neural network. However, data dependent control flow like the data routing for sparse mixture of experts cannot be described with a standard static computation graph upfront [57].

There are limits on what can be decided about a program using static analysis, for example the halting problem, or more generally Rice's Theorem [58]. It is unclear how relevant these limits are for guarantees about accelerator workloads. If the compute graph is dynamically constructed and checked for guarantees this is not an issue because branch conditions do not need to be represented (instead logging which mathematical operations were actually performed).

Existing libraries like PyTorch or CUDA graphs could be modified to provide compute graph declaration functionality, especially static compute graph declarations that these libraries already support. Alternatively, a description language specifically for flexHEG compute graphs could be developed along with tools to translate accelerator kernels into this language. This approach could use solutions from existing programming languages that have already addressed similar challenges in describing mathematical operations on data (and the problem is simpler than building a programming language because how the operation is performed does not need to be described)

Accelerator workloads typically involve simple mathematical operations on large tensors, making it natural for the flexHEG compute graph description to be based on atomic operations applied to blocks of memory. One approach would be to allow developers to describe these atomic operations using general-purpose code (such as C or similar languages), though this would require careful consideration of security implications. Alternatively, a library of pre-defined, checkable atomic operations could be developed. The number of operations needed is relatively limited since accelerators are specialized hardware - for example Google's TPUs which primarily support tensor operations, arithmetic, and a few transcendental functions [59]. Each operation could include type descriptions specifying input and output formats (such as tensor dimensions and numeric representation).

Either the CPU or accelerator could be responsible for communicating declarations about the intended workload to the Guarantee Processor, either beforehand with static declarations, or as kernels are sent to the accelerator with dynamic declaration.

To increase trust in the declared compute graph, the guarantee processor can ideally independently verify that the operations are performed as claimed. If the accelerator is trusted enough to follow the input kernel instructions, the kernels could directly be parsed and accepted as truth for atomic computations (assuming the guarantee processor can robustly monitor which kernel instructions are received by the accelerator). Alternatively, a general-purpose way to verify each atomic computation is to use the guarantee processor to randomly re-compute a portion of each result and verify that it matches the value that was returned. Data transfer between accelerators or nodes could be checked by the Interlock using authenticated cryptography, as described in [Encrypted Cluster Formation](#).

The timing of these declarations and checks would need to be carefully designed to avoid vulnerabilities from mistimed checks, or inefficiencies imposed on the model developer by prohibiting types of asynchronous operations that are difficult to check. Additionally, any non-accessible or untrusted internal accelerator state that influences the computation result would make performing these checks challenging.

Because accelerators operate in a highly parallelized, single instruction multiple data mode, it may be possible to store the description of the kernels with very little data. As a rough estimate, each accelerator may process 10 kernels per second, and each kernel can be described with 1KB, leading to 10KB/s data created per accelerator. Unfortunately, locally storing a copy of the whole compute graph on each flexHEG system does not scale well with cluster size. For example, if there were 100k accelerators casually contributing to one workload, each would contribute 10KB/s to the overall instruction log, causing the compute graph to grow by 1GB each second. With 1M accelerators, this would be 10GB/s, which exceeds SSD write speeds, and would almost instantly fill up the RAM available on a single device.

For most types of guarantee checks on large clusters, the compute graph would have to be analysed in a distributed way. How this is implemented would depend on the type of guarantee, but might involve computations that could more compactly summarize the guarantee-relevant parts of the compute graph. These distributed guarantee checks could potentially be done periodically, for example each time a snapshot of the weights is saved.

For a few specific types of guarantees, the necessary information from the compute graph can be summarized efficiently on the fly, so that each flexHEG system can locally store the relevant global state even with large clusters. In [FLOP Counting](#), we describe one way to implement this by locally keeping a running tally of the FLOPs performed by every accelerator in the cluster that could have casually influenced the data.

## Framework-Based Verification

Another approach to verification is to develop a framework that constrains how developers can interact with the hardware. Such frameworks can enable verification in two ways:
1. Structural Constraints: The framework's API design could only provide limited functionality, or even enforce guarantees directly:
    - The framework could simply not provide APIs for certain operations, making some non-compliant workloads impossible to implement (e.g. requiring static declarations of computations, therefore preventing data-dependent looping)
    - Resource allocation could be handled entirely by the framework, ensuring compliance with hardware usage rules

2. Built-in Verification: When developers must work through framework APIs, verification logic can be built directly into these interfaces:
    - Developers might need to declare model architecture parameters upfront, making properties like model size or layer count trivial to verify

Simply not providing the ability to execute non-compliant workloads would avoid false positives and make guarantee checking more robust, though it requires developers to adapt their workflows to the framework's constraints. Additionally, this framework would ideally be compatible with multiple kinds of hardware platforms. Instead of starting from scratch, it might be easier to adapt existing open source tools like PyTorch to build in guarantees, although this comes with the risk of vulnerabilities in code that was not originally intended to be secure.

## Guarantee Example: Multi-Accelerator FLOP Counting

This section provides more specific implementation details than the general-purpose workload monitoring strategies discussed in the previous section. The number of operations used to create an AI model serves as a proxy for that model's capabilities [Leonie] and has become an important component of European and American AI regulation [EU AI act, diffusion rule]. While FLOPs can be tallied through classical accounting approaches—such as requiring developers to declare their number of accelerators and document their high-level tasks over time [60]—these methods have potential vulnerabilities. Standard accounting practices might fail to detect deliberate misrepresentation of FLOPs. For example, a developer that has their own data center could falsely claim that 80% of their compute is dedicated to inference while 20% handles training, when in reality the entire facility is being used for a large training run. In scenarios where compute resources must be tightly networked, such misrepresentation could be detectable by cloud providers through node connection monitoring or hardware-backed network reporting (as discussed in Encrypted Cluster Formation). However, developers could still circumvent restrictions by claiming to run several unrelated training runs sequentially that are actually components of a single large training run.

As a more robust option, the Interlock could be used to tally all inputs that could have causally contributed to output data. Below, we sketch a potential implementation of this for multi-accelerator workloads. We assume that the Interlock is able to keep an up-to-date log of the FLOPs performed by the local accelerator after it is powered on. This could be done using performance counters, by scanning executed kernel code, approximating based on power usage, or with other methods.

The number of FLOPs that could have contributed to an output is the sum of FLOPs performed on the local accelerator and the FLOPs performed by other accelerators that generated data that was used by the local accelerator. To prevent FLOP "laundering" we must also account for data that is entered into the cluster system from outside sources.

Typical accelerator workloads involve external data, which could be static data used for training, or interactive data from users. Without protections, results from a previous or parallel workload could be hidden in the "external data", falsifying the total FLOP count. This can be protected against in a few ways: 1) the external data could require an audited explanation, or 2) could be delayed to make it inconvenient to bridge parallel or sequential workloads. Small amounts of data could be allowed to enter the system without an explanation or significant delay, but the amount of unexplained data would have to be logged so that it does not accumulate to a large amount over time.

To provide an audited explanation of external data, data that enters the system could be hashed. The hash and amount of the data could be included as a new entry in the FLOP log. Later, it could be verified that this hash matches a hash provided by an auditor attesting that they inspected that specific data to check that it was human-generated (perhaps by randomly sampling fractions of the data and verifying sources).

To enforce a large delay, a similar mechanism could be used. Instead of an auditor claiming that the data is authentic, they could simply claim to have seen that hash a certain amount of time before (or instead of an auditor, the presence of that hash in the history of a public blockchain could be used). For preventing the sequential combination of large workloads, this delay might have to be on the order of months. E.g. if the flexHEG system allows workloads up to $10^{26}$ FLOPs and then halts, a delay of 3 months would substantially slow attempts to restart a "new" workload from a checkpoint of the previous workload. For this example, to reach $10^{27}$ FLOPs sequentially would require 10 restarts, taking at least 30 months. For preventing unauthorized parallel combination of workloads, potentially even seconds of latency could be sufficient to meaningfully slow circumvention attempts [61]. This delay could be added by the Interlock itself if it has a large enough buffer to store all external data for a few seconds.

The amount of information that needs to be tracked about each accelerator is small, so this calculation can be done efficiently on-the-fly without requiring a whole-cluster summarization step that could be complicated, especially if it has to deal with accelerators unexpectedly going offline. For each source, 1 bit could specify whether it is unknown data or another accelerator. Each accelerator session ID could be a 16 byte UUID, with another 12 bytes providing more than enough space to store the local FLOPs as an integer. If it is unknown data, 32 bytes could store the hash.  Overall, this would be roughly 30MB to store the FLOP log for 1M unique sources (e.g. a very large cluster). An additional 30MB is not significant compared to the GB-scale data that is typically transmitted between accelerators. If necessary, some accelerators could be excluded from the local log depending on the network topology (e.g. intermediate nodes could summarize groups of FLOP counts).

> On-the-fly distributed FLOP counting:
>
> 1. At initialization, each accelerator is power-cycled (turned off then back on) to delete any stored data[11].
> 2. The Interlock generates a new session ID for itself and sets the FLOP counter for this session ID to 0.
> 3. Data sent to other accelerators is sent with a prefix that contains the Interlock's entire FLOP log.
> 4. When an Interlock parses input data, it reads the prefix, adds any new sources to its local log, and for each source it updates the FLOP value if the new value is higher

The Interlock needs to store the FLOP count of every source individually and not just a single running tally to avoid double counting[12] . With this data flow, each Interlock doesn't necessarily have the current FLOP count of every other accelerator, but it does have the highest FLOP count that could have casually influenced the local data[13].

Some care has to be taken if the output data is streamed, and is not all staged in the Interlock at the time the prefix FLOP log is sent. This is because the count could be misrepresented if additional FLOPs could influence the output after the prefix has been generated but before all the data has been sent. To prevent this, the prefix log could be modified to include an additional buffer accounting for additional local FLOPs that might occur during streaming. If the local count exceeds this pre-specified buffer, then the output stream has to be halted. Any incoming data could be staged in the Interlock until the output stream has completed so it couldn't influence the output stream. This additional buffer quantity added to local FLOPs would only introduce a fixed error and not an accumulating error, because the FLOP value is tracked as the maximum and not summed on other accelerators.

Data that is exported from the workload can include a signed certificate describing the detailed FLOP log. The detailed log is useful so that multiple chunks of data can be combined later without double counting the FLOPs (because the logs show overlapping source IDs, the FLOPs can be summed without double counting, and then combined into a single value as necessary).

---

[11] It should be turned off for long enough so that data in RAM is not retained
[12] E.g. doing a computation with 2000 FLOPs, sending it to another accelerator that does 1000 FLOPs, then that other accelerator sends it back, saying that it took 3000 FLOPs. With a single FLOP counter, it would partially double count this as 5000 FLOPs, instead of the true 3000 FLOPs
[13] With the potential exception of work done towards messages that are not sent, e.g., failed attempts at generating synthetic reasoning traces.

# Guarantee Update Process

For the guarantee logic to be flexible, it must be possible to update it on deployed devices. However, to prevent attackers from disabling the guarantees, the guarantee processor should only accept updates that were authorized by the appropriate parties. On consumer devices, this authorized update process is typically implemented using public key cryptography, where the new firmware must be cryptographically signed using the manufacturer's private key for it to be accepted by the device [62]. A similar solution can be used for flexHEG, with a few modifications:

- The update process could require that the update (or lack of update) be multilaterally approved by k of n parties, so that no single actor could unilaterally impose excessive restrictions (as discussed for an international context in [Part III](#)). The pseudocode in algorithm 1 shows how multi-party update approval could be implemented.
- Different amounts of authorization (e.g. number of stakeholder signatures) could be required for different types of updates. For example, whole-system updates might require a supermajority, while changes to a parameter file within a specified range of values could require a smaller majority approval.
- An update could change how future updates are accepted. This could change which parties are able to approve updates or how many votes are needed for different types of updates.
- Signatures will ideally use quantum resistant public key cryptography [63] so that security is not broken by advances in quantum computing (which could plausibly be sped up with AI advancements).

```
def verify_firmware_update(firmware_binary: bytes,
            signatures: list[tuple[str, bytes]],  # [(signer_id, signature)]
            authorized_signers: dict[str, bytes],  # {signer_id: public_key}
            k: int,  # Required number of valid signatures
             ) -> bool:
   # Calculate firmware hash
   firmware_hash = hashlib.sha256(firmware_binary).digest()

   # Count valid signatures from unique signers
   valid_signers = set()
   for signer_id, signature in signatures:
      # Skip if not an authorized signer
      if signer_id not in authorized_signers:
         continue

      # Skip duplicate signers
      if signer_id in valid_signers:
         continue

      # Verify signature using signer's public key
      public_key = authorized_signers[signer_id]
      if verify_signature(firmware_hash, signature, public_key):
         valid_signers.add(signer_id)

      # Return early if we have enough valid signatures
      if len(valid_signers) >= k:
         return True

   return False
```

Algorithm 1: Pseudocode for k of n update authorization process. Updates could themselves change the future authorization process by either changing the authorization logic, or the list specifying expected signers. Authorization logic can include more (or less) complex rules if necessary.

It may be useful to be able to provide evidence about which updates were and were not signed by particular stakeholders. Updates that were signed are trivial to prove (by showing the signature), but showing that an update was not signed would probably require hardware-backed evidence from all devices that contain the private key.

The simplest update deployment strategy is to publish cryptographically signed updates and allow device operators to download and install them if/when they choose. If the flexHEG system is only being used to provide credible attestation about how accelerators are used, voluntary updates would be sufficient (because the attestation would communicate which set of guarantees it is for). However, if the flexHEG system needs to guarantee compliance with the updated rules, voluntary updates would be insufficient, because device operators could simply not install the updated guarantees.

An alternative approach that could be used to enforce updates is to require that accelerators periodically receive a license in order to continue operating [11], [12], [13], and for each license to specify a minimum allowed version number. The guarantee processor could prevent usage of the accelerator if it has operated for too long (as measured by a secure real-time clock or cycle counter[14]) without receiving a license, or if the minimum version number is greater than the current version number. The licenses and updates can be delivered by any channel, so no internet connection is required for this approach (which could be important for chips that are air-gapped for security).

The length of time that a license is valid affects the latency at which new updates must be loaded (e.g., if devices need a new license every three months, they could ignore an update for that much time). This pushes in the direction of making the license renewal time shorter so that guarantee changes are not overly delayed. However, with very frequent license renewals, the burden on operators becomes more significant. One approach that could be used to resolve this is to use data diodes [64], which could be used to transmit licenses without any risk of data leaving via the same path.

It may be useful to be able to commit to not deploying certain types of updates in the future. To make specific commitments would require unmodifiable structure in the guarantee logic. Being unable to deploy some types of updates would (by design) reduce the ability to flexibly enforce future rules. One way to implement this is to provide a minimal guarantee version (See discussion of "baseline rulesets" in Parts I and III) that partially throttles accelerator performance, but cannot be revoked. Choosing a minimal level is not simple, because too high would limit the potential for governance via the mechanism, and too low would be almost the same as not having a minimal version at all. A major downside of providing a fallback version or other restrictions on future rules is that if security vulnerabilities are found in the fallback version they would not be possible to patch (without cooperation from the users).

Choices for the guarantee update process in the context of international agreements are analyzed in greater detail in Part III of this report series.

# Future Work

As discussed throughout this report, there are several concrete paths to developing flexHEG capabilities. In this section, we outline promising directions for a major third-party R&D effort.

---

[14] If using a cycle counter, there is a risk that the operator secretly does not use the chip to save up usage time. This risk could potentially be mitigated by requiring usage reports.

Assistance from accelerator designers could dramatically speed up this work by providing information for seamless integration or by directly contributing to the R&D efforts.

The NIC appears to be the most promising component that third parties could modify to enable secure, trusted, and flexible guarantees. IP blocks, code, and algorithms originally developed for a NIC-based solution could later be adapted for Guarantee Processors in other locations if necessary. One caveat to this is that deploying a flexHEG-enabled component to major data centers would require substantial cooperation from the entities that normally manage those components (e.g., accelerator designers or data center operators). Performance-critical components like the accelerator chip, HBM, or NIC would likely require more extensive collaboration compared to more peripheral components like the BMC or DC-SCM. If minimal deployment support is expected, a pragmatic approach would focus R&D on minimally disruptive solutions, though this trades off against security and capabilities.

Among the guarantee-checking strategies described earlier (in General-Purpose Guarantees), developing dynamic compute graph logging with automated guarantee checking would be the most significant breakthrough. Dynamic compute graph checks would offer greater privacy preservation than manual auditing, more robust verification than opportunistic classification, and more flexibility than structured hardware access. However, it requires significant research before deployment, making it well-suited for a major research project. For these reasons we recommend focusing ambitious research efforts on dynamic compute graph logging, while pursuing faster deployments using alternative approaches.

Below, we outline a roadmap for how a major third-party effort ($10M-$100M) could develop a working version of flexHEG that is compatible with leading accelerators. This roadmap focuses on developing a tamper-evident enclosure, NIC-based interlock, and automated guarantee checking using dynamically logged compute graphs.

We divide this roadmap into four mostly independent projects, followed by integration work to combine the results into a frontier deployment. Each project depends on the success of previous projects, but many parts can be developed in parallel. It may make sense to start each of these projects at a modest size and then quickly scale projects that need more attention in order to be viable, or that could best make use of additional effort.

These project suggestions are intended as initial recommendations, with the expectation of changing course as more information is gathered (where possible trying to avoid changes that would affect high level goals or the other subprojects). If a major roadblock is discovered, it may make sense to switch to an easier target design, for example by using the same Guarantee Processor and Secure Enclosure, but using opportunistic classification instead of dynamic compute graph logging. If there is enough research capacity, it may make sense to reduce technical risks by exploring these alternative designs in parallel.

# Project 1: Tamper-Evident Secure Enclosure

**Summary:** Develop a best-effort solution for tamper-evident physical security for AI accelerators that could be deployed within 1-3 years.

**Related Sections:** Secure Enclosure

**Deliverables:**

- Specific recommendations for applying existing technology to create a tamper-evident enclosure
- Evidence that the proposed design is secure and not overly disruptive to data center operations, ideally with red-team testing

**Primary skill sets:** Hardware Engineering, Hardware Security, Data center operations

**Time Estimate:** 2-10 person-years to adapt existing tamper evidence methods to data center accelerators (and more if there are promising directions for advancing the frontier of hardware security)

**Details:** The aim of this project is to provide a minimal level of tamper evidence that could be retrofitted to existing data centers to enable credible guarantees. Tamper resistance would be valuable to develop simultaneously, though it is a lower priority as it will likely be developed commercially to protect model weights.

Results should be actionable within 1-3 years, so reliance on unproven technology should be minimized. If promising approaches to highly secure tamper-responsive enclosures emerge that could defend against overt nation-state adversaries, this work could be separated into a distinct project (governments with access to classified information on hardware security might be better positioned to assess feasibility).

The project should begin by thoroughly investigating current physical security practices and maintenance requirements for AI data centers. Next, the team should survey options for tamper evidence, potentially including tamper-evident seals, locks, and camera systems. Insights could be drawn from technologies used in nuclear non-proliferation treaties (noting that some seals have reportedly been compromised [65]).

Promising options should be tested under conditions approximating real-world deployment, with maintenance protocols also addressed. This may include strategies to minimize maintenance requirements or structured access protocols to prevent maintenance operations from being used to disguise tampering. Ideally, the complete proposed solution would undergo external red-team security assessment and usability testing.

## Project 2: Trustworthy Guarantee Processor

**Summary:** Analyze existing NICs for use as a Guarantee Processor, and develop a trustworthy, open-source Guarantee Processor IP block.

**Related Sections:** [Guarantee Processor](#)

**Requires:** Physical security provided by Project 1

**Primary Skill Sets:** Chip design, chip analysis

**Deliverables:**

- Recommendation on ways to use existing processors in a secure and trusted way
- Open-source design for a Guarantee Processor IP block that could be integrated into other hardware

**Time Estimate:**

- 2-4 person-years to analyze existing options
- 10-40 person-years to design the IP block

**Details:** As discussed previously ([Guarantee Processor](#)), a Guarantee Processor should ideally incorporate the following features to ensure security, auditability, and the ability to execute logic required for Projects 3 and 4:

- FPGA and/or general-purpose processor capability
- Sufficient processing power to keep up with 1 of the 192 Streaming Multiprocessors on a Blackwell chip (though less could be workable)
- Secure boot with rollback protection
- Hardware for public key cryptography (ideally quantum-secure)
- Lockstep processor execution capability for security
- Secure random number generation
- Glitch protection
- Sufficient non-volatile memory to store months of logs locally
- Secure non-volatile memory to detect modifications of partially untrusted non-volatile memory

Additionally, the following are NIC-specific requirements:

- Hardware for AES-GCM (ideally 800Gb/s or 1600Gb/s, or a clear path to scaling to this rate)—alternatively, the ability to control external AES-GCM hardware in a secure and trustworthy way

- Buffer on the data path (large enough for the Guarantee Processor to read a small fraction of data as it passes through)
- DMA capability to read accelerator memory

The project should first assess the security, auditability, and performance of existing high-performance NICs, especially those already deployed with accelerators, such as NVIDIA's ConnectX and Amazon's Nitro. If possible, plans should be developed to quickly repurpose these NICs for flexHEG applications with support from the NIC designers.

Next, high-level plans should be developed for integrating a Guarantee Processor IP block into future leading NIC designs. This could begin with a thorough survey of existing open-source IP blocks that could serve as a foundation, such as OpenTitan or similar projects.

The team would then design an IP block with the features required for flexHEG. Beyond the performance and security requirements listed above, the IP block should be trustworthy to third parties not involved in its design or manufacturing (assuming they can conduct random hardware inspections).

The IP block should be manufactured and tested to verify readiness for real-world deployment. If possible, this testing should include security red-teaming and trustworthiness auditing.

Ideally, this IP block would be designed to be compatible with future NICs, other accelerator components, or as an additional chiplet on accelerator boards. The IP block should be open-sourced to enable auditing and integration by multiple manufacturers.

## Project 3: Accurate Compute Graph

**Summary:** Implement dynamic compute graph logging, including hardware-backed verification methods.

**Related Sections:** Compute Graph Declaration

**Requires:** Trustworthy Guarantee Processor provided by Project 2

**Primary skill sets:** Applied Mathematics, Firmware Engineering

**Prototyping platform:** Reprogrammable logic on standard smartNIC (like ConnectX) and/or CPU

**Deliverables:**

- Language to concretely and concisely describe multi-accelerator workloads, including computations performed on memory blocks and data movement between accelerators

- Robust methods to verify accuracy of each atomic operation in a workload (e.g., random recomputation for calculations, in-line cryptography or memory log comparison for data transfers)

**Time Estimate:**

- ~20 person-years to create an extensible description language and checking system
- ~4 person-years to develop tooling to automatically translate accelerator kernels to the workload description language

**Details:** The project should begin with empirical work to understand which operations must be supported and the precise timing of kernel execution and HBM data modifications across multiple accelerator types and relevant workloads. This data could be collected by running typical workloads and logging detailed information about accelerator kernels and HBM state.

The team would then develop a representation for workloads that is specific enough to verify hardware execution of claimed operations, flexible enough to accommodate potential uses, but not so flexible that analysis of these logs is intractable. This could build on structure from existing frameworks like PyTorch [66], JAX [67], TensorFlow [68], StableHLO [69], Aesara [70], MIL [71], IREE [72], ONNX [73], LiteRT [74], TOSA [75] or WebNN [76].

The next step would be creating a protocol for the CPU or accelerator to make real-time claims about how each memory block will be updated. These declarations could be shared with the flexHEG NIC using DMA. This protocol could be prototyped using user-programmable logic on an existing smartNIC or FPGA attached to a leading accelerator (with security and auditability requirements skipped until later).

The team would then design verification checks for each operation to confirm it occurred as claimed. With a NIC Interlock, network communication could be directly verified. DMA could be used to read portions of HBM data, enabling recomputation of a sample of output values. Important threat modeling questions to address include:

- If DMA could be falsified but network communications are accurate, how difficult would it be for an attacker to falsify the overall compute graph?
- How does this change if an attacker is willing to dynamically modify a portion of their workload to appear compliant?
- Can checks be designed so that the complexity for attackers increases with each operation, rather than requiring a single patch of their data?
- Could supplementary measurements like accelerator kernels, power usage, or timing corroborate claims?
- Are there ways to increase confidence in PCIe reads of HBM?
- Would placing the Guarantee Processor on HBM or the accelerator chip (instead of the NIC) significantly improve the defense model?

Tools should be developed to translate accelerator kernels to properly formatted workload descriptions, potentially leveraging existing software like StableHLO.

Finally, the compute graph construction process should be tested with typical workloads to verify functionality.

## Project 4: Automated Guarantee Checking

**Summary:** Develop classifiers to robustly verify policy-relevant guarantees using computation graphs from multiple Guarantee Processors.

**Related Sections:** Compute Graph Declaration, FlexHEG Part I Appendix B: Governance Mechanisms Enabled by FlexHEGs

**Requires:** Accurate compute graph provided by Project 3, trustworthy Guarantee Processor provided by Project 2 (or another similarly secure and trustworthy processor)

**Primary Skill Set:** Computer Science / Mathematics

**Prototyping platform:** Mathematical modeling and/or software (on any platform)

**Deliverable:** Algorithms that can process workload logs from multiple Guarantee Processors and robustly classify whether the combined workload meets policy-relevant guarantees.

**Time Estimate:**

- 4 person-years for basic guarantees
- 10+ person-years for exploration of more sophisticated guarantees

**Details:** The objective of this project is to automatically process Guarantee Processor logs describing compute graphs to determine whether workloads meet policy-relevant guarantees. If the workload log format is not finalized when this project begins (likely if Project 3 starts at the same time), work on automated checks could begin using similar specifications, such as PyTorch compute graphs.

The first step would be creating a protocol to assemble or process compute graphs from multiple Guarantee Processors in a scalable way that can tolerate potential hardware failures.

Next, the team would develop robust guarantee checks, prioritizing development based on policy relevance and technical feasibility, likely focusing on simpler checks to start.

Development could begin with classifiers for lower-level properties (which should be relatively straightforward once the compute graph is constructed), such as:

- Number of linked accelerators
- Number of FLOPs that contributed to an output

The team could then develop higher-level guarantee checks, such as:

- Whether gradient descent is occurring
- Whether reinforcement learning is being performed
- The length of particular LLM chains of thought
- Whether evaluations were run properly

Red-team testing should be used to assess the robustness of the automated checks developed. If there is time, the team should also analyze robustness against falsification of some operations in the compute graph.

## Integration and Deployment

After completing the four projects, they should be combined into a full flexHEG system. Specifically:

- The Guarantee Processor (and linked accelerator) from Project 2 should be protected by the Secure Enclosure from Project 1
- The compute graph construction process from Project 3 should be executable on the Guarantee Processor from Project 2
- The automated guarantee checking from Project 4 should be able to analyze compute graphs from Project 3 (and ideally run on the Guarantee Processor from Project 2, though a dedicated device could be used if necessary)
- Typical frontier workloads should run efficiently on the combined system with minimal setup requirements

Follow-up work should include security red-teaming, user experience testing, and developing standards to credibly assess trustworthiness and security of solutions from other parties. Standardizing hardware and logical interfaces would improve compatibility with novel accelerators.

The work in these projects is non-trivial but likely feasible for a dedicated flexHEG accelerator project. The primary uncertainty is ease of integration on short notice with modern hardware and frontier workloads. If integration is straightforward, deployment could occur within 2-4 years, either with a firmware update or by retrofitting flexHEG components. However, if integration is more challenging, deployment would have to wait for accelerator designers to include it themselves and release updated accelerators. An upside of deeply integrated flexHEG is that it could be significantly more secure and have access to more reliable data.

Uncertainty about integration could be de-risked with more extensive investigation of:

- Maintenance frequency for AI accelerators and realistic reduction potential
- Specific timing of accelerator instructions and memory changes to better understand the feasibility of the proposed claim-and-verify architecture for dynamic compute graph logging

A successful third-party flexHEG project could demonstrate feasibility, resolve technical uncertainties (especially around automated guarantees), and serve as a foundation for ongoing industry development. Findings could inform firmware updates providing partial flexHEG functionality on already-deployed accelerators.

Given the strategic importance of AI, the significant lead times required for hardware development, and the growing need for privacy-preserving verification mechanisms, accelerating the development of flexHEG capabilities now is important so that these solutions are ready if they are needed.

Flexible Hardware-Enabled Guarantees | Part I | **Part II** | Part III[65] R. G. Johnston, A. R. E. Garcia, and A. N. Pacheco, "Efficacy of Tamper-Indicating Devices".
[66] "PyTorch." Accessed: Apr. 14, 2025. [Online]. Available: https://pytorch.org/
[67] *jax-ml/jax*. (Apr. 14, 2025). Python. jax-ml. Accessed: Apr. 14, 2025. [Online]. Available: https://github.com/jax-ml/jax
[68] "TensorFlow." Accessed: Apr. 14, 2025. [Online]. Available: https://www.tensorflow.org/
[69] *openxla/stablehlo*. (Apr. 11, 2025). MLIR. OpenXLA. Accessed: Apr. 14, 2025. [Online]. Available: https://github.com/openxla/stablehlo
[70] B. T Willard *et al.*, *Aesara*. (Jun. 2023). Python. Accessed: Apr. 14, 2025. [Online]. Available: https://github.com/aesara-devs/aesara
[71] "Model Intermediate Language — Guide to Core ML Tools." Accessed: Apr. 14, 2025. [Online]. Available: https://apple.github.io/coremltools/docs-guides/source/model-intermediate-language.html
[72] The IREE Authors, *IREE*. (Sep. 2019). C++. Accessed: Apr. 14, 2025. [Online]. Available: https://github.com/iree-org/iree
[73] *onnx/onnx*. (Apr. 14, 2025). Python. Open Neural Network Exchange. Accessed: Apr. 14, 2025. [Online]. Available: https://github.com/onnx/onnx
[74] *google-ai-edge/LiteRT*. (Apr. 14, 2025). C++. google-ai-edge. Accessed: Apr. 14, 2025. [Online]. Available: https://github.com/google-ai-edge/LiteRT
[75] "Developer Resources," Linaro. Accessed: Apr. 14, 2025. [Online]. Available: https://www.mlplatform.org/tosa/
[76] stevewhims, "WebNN Overview." Accessed: Apr. 14, 2025. [Online]. Available: https://learn.microsoft.com/en-us/windows/ai/directml/webnn-overview
[77] "Supermicro NVIDIA GB200 NVL72 SuperCluster".
[78] "Preventing AI Chip Smuggling to China | CNAS." Accessed: Apr. 14, 2025. [Online]. Available: https://www.cnas.org/publications/reports/preventing-ai-chip-smuggling-to-china
[79] "Data protection in Amazon EC2 - Amazon Elastic Compute Cloud." Accessed: Apr. 14, 2025. [Online]. Available: https://docs.aws.amazon.com/AWSEC2/latest/UserGuide/data-protection.html#encryption-transit
[80] E. AI, "Machine Learning Trends," Epoch AI. Accessed: Apr. 14, 2025. [Online]. Available: https://epoch.ai/trends
[81] A. Grattafiori *et al.*, "The Llama 3 Herd of Models," Nov. 23, 2024, *arXiv*: arXiv:2407.21783. doi: 10.48550/arXiv.2407.21783.
[82] "Distributed communication package - torch.distributed — PyTorch 2.6 documentation." Accessed: Apr. 14, 2025. [Online]. Available: https://pytorch.org/docs/stable/distributed.html
[83] "Introducing Trillium, sixth-generation TPUs," Google Cloud Blog. Accessed: Apr. 14, 2025. [Online]. Available: https://cloud.google.com/blog/products/compute/introducing-trillium-6th-gen-tpus
[84] "The Ultra-Scale Playbook - a Hugging Face Space by nanotron." Accessed: Apr. 14, 2025. [Online]. Available: https://huggingface.co/spaces/nanotron/ultrascale-playbook50

# Appendix A: Background on Frontier AI Development and Hardware

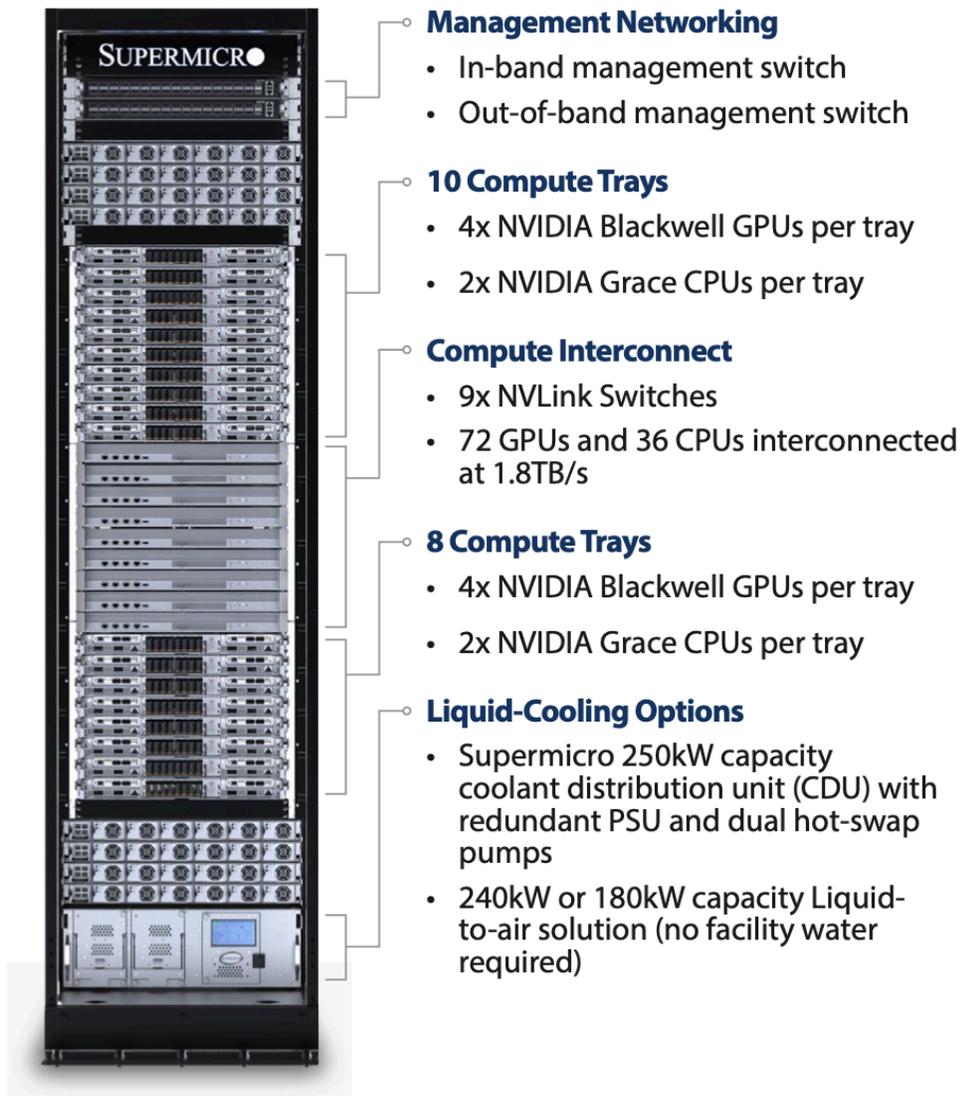

Figure 1: NVL72 rack, with 72 Blackwell accelerators, direct-to-chip liquid cooling, and 9 NVLink switches [77].

This section provides a brief review of how frontier AI workloads are run on specialized accelerators in large data centers. FlexHEG designs that are minimally disruptive to these existing systems could be integrated more quickly.

Most accelerators that are currently in use are designed by NVIDIA and manufactured by TSMC, although AMD, Google, Amazon, Intel and Huawei have competing accelerator designs which some major AI developers claim are their primary drivers for AI training. Chinese AI firms currently use handicapped chips that comply with export controls, and likely also use cloud compute and smuggled chips [78]. Chinese manufacturers are attempting to develop semiconductor processes that are competitive with TSMC, although it is unclear how long this might take.

Hyperscalers (Microsoft, Amazon, Google, etc.) have massive data centers for AI training and inference, and are currently building many more. According to Epoch AI, the total installed NVIDIA compute is increasing approximately 2.3x per year (Figure 2). Semianalysis predicts that NVIDIA Blackwell GPUs in an NVL72 configuration will be a major part of this ongoing buildout . To briefly summarize the Semianalysis report on NVL72, the layout essentially consists of one or two datacenter racks with direct-to-chip liquid cooling, enabling higher compute/power density compared to traditional air cooling. Each compute tray in the rack contains two GB200s (Figure 3), each comprising two Blackwell GPUs, a Grace CPU, two CX-7 Network Interface Cards (NICs), and outgoing NVLink connections. These NVLink connections route to NVLink switches housed in a separate switch tray, providing high-bandwidth communication between every GPU in the NVL72. The CX-7 NICs connect to the backend datacenter network (either InfiniBand or Ethernet) via optical cables, allowing multiple NVL72 racks to communicate with each other. A separate frontend network with a different set of network switches handles communication with the internet and networked memory (e.g., weight snapshots). Additionally, an out-of-band management network supports physical monitoring via a module called the baseboard management controller (BMC).

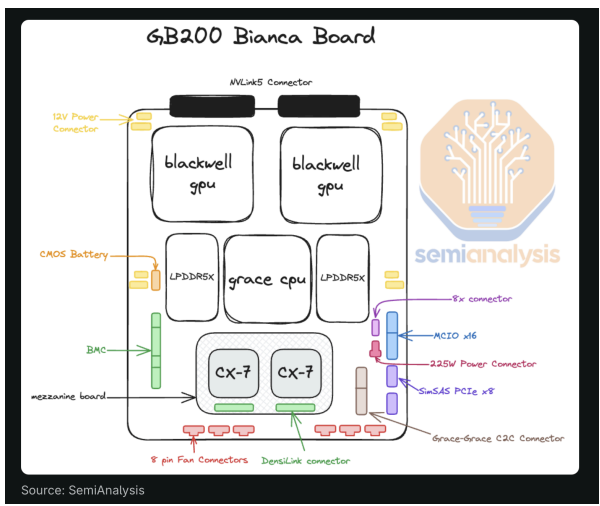
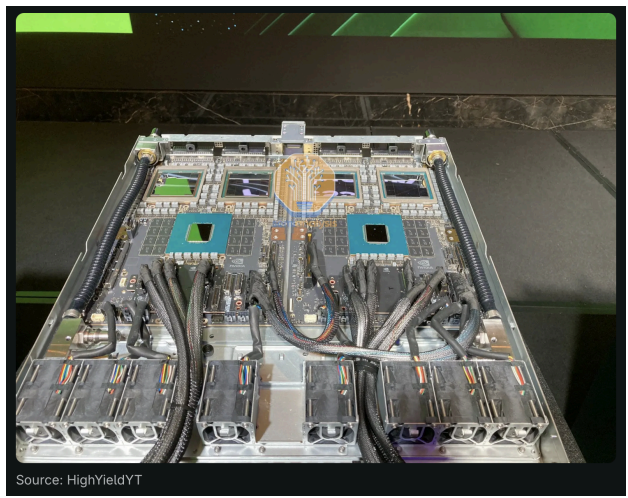

Figure 3: GB200 Bianca board layout (left, source: Semianalysis [52]) and NVL72 tray with two Bianca boards (right, source: HighYieldYT).

Modern data centers increasingly use hardware security features and encrypted communication to protect sensitive workloads. For example, AWS encrypts network traffic between EC2 instances using TLS [79]. Some hardware cryptographic modules are designed to be resistant to physical tampering and side channel attacks, and the FIPS-140 standard aims to certify the security of these crypto modules [38]. Intel, AMD, and ARM offer CPUs that can support a Trusted Execution Environment (TEE) [15], which provides hardware-backed assurances that the computing environment is configured as claimed. Similarly, NVIDIA has developed confidential computing capabilities for their GPUs, initially supporting single-GPU workloads with plans to extend this to multi-GPU scenarios [50].

Frontier AI development in 2020-2024 has been dominated by massive pre training runs, where LLMs using the transformer architecture are trained on huge amounts of text [80]. As a typical example of frontier development, Meta's Llama 3 paper [81] describes their approach for training a dense 405B parameter model on 16,000 H100s. They state it "uses a standard, dense Transformer architecture" and appears to employ a fairly typical training procedure, although this is difficult to confirm since many frontier labs do not publish their training methodology. Notably, their training run experienced 466 interruptions over 54 days of training, and "78% of the unexpected interruptions [were] attributed to confirmed hardware issues." Their workload parallelism configuration was chosen based on network bandwidth and latency between GPUs, and they attempted to overlap computation and communication. Meta and several other frontier AI developers use PyTorch for defining their AI training workload [66]. Pytorch allows developers to define a model, convert it to a computation graph, use automatic differentiation for the gradient calculation, and distribute the entire workload across many GPUs. The work done by individual GPUs in this distributed workload is described by CUDA kernels, which proprietary NVIDIA drivers convert to machine instructions. Pytorch distributed communication operations can use NVIDIA's NCCL backend which can optimize GPU communication patterns [82].

Google trains their models using their proprietary TPU accelerators [83] (rather than Nvidia GPUs), and JAX [67] (rather than Pytorch). Their book on model scaling explains how to develop distributed training code that is optimized for their hardware [59]. Similarly, Hugging Face has a book on how to optimize distributed training to reduce compute, memory, and network bottlenecks [84].

As another example of frontier model development, the Chinese company Deepseek published on the training process for their Deepseek V3 model, including several model and infrastructure optimizations to overcome compute limitations [57]. Due to export controls, they were unable to purchase H100s, and instead used H800s which have about half the NVLink bandwidth. The training run used 2048 H800s over 56 days (about 8x fewer GPUs than were used for LLama 3 405B). They used sparse Mixture of Experts, which selects a subset of the network to route each token to at each layer. With this sparse architecture, only 37B parameters out of 671B total parameters are used for inference per token. They also use a mixed precision framework and carefully overlap computations with communication. Information on the software stack used to develop Deepseek V3 is not currently publicly available.

OpenAI recently announced their o1 and o3 models, which achieved impressive benchmarks by scaling inference-time compute using chain of thought. Details about o1 and o3 architecture and training process have not been published, although there is speculation about synthetic data and reinforcement learning. It is currently unclear how much this will change the allocation of compute for AI development, and how important it will be to be able to make guarantees about workloads other than pre-training.

# Appendix B: Threat Models

FlexHEG systems aim to provide hardware-backed guarantees for AI accelerators, but these mechanisms face potential threats from actors attempting to circumvent the guarantees or misuse the systems. This appendix analyzes threat models relevant to different flexHEG deployment scenarios.

Several types of attacks could be used for these purposes, including:
1. Exploiting firmware vulnerabilities (e.g., buffer overflow attacks)
2. Exploiting mispecification in the guarantee logic (e.g., causing workloads to be inaccurately classified)
3. Physical tampering with components that enforce restrictions on chip usage, potentially using Fault Analysis tools to image the device and make targeted modifications (e.g., resetting a FLOP counter)
4. Leveraging backdoors inserted during design or manufacturing processes
5. Making secret modifications to cluster configurations (e.g., adding bridge devices to combine supposedly separate clusters)
6. Attacking the infrastructure used to authorize flexHEG devices (e.g., stealing authorization keys)

The plausibility and impact of these attack vectors vary significantly based on the use case and the adversary's capabilities and motivations. We analyze four scenarios: domestic oversight, international treaty verification, international treaty enforcement, defending AI system IP.

## Domestic Oversight

In the context of domestic regulation, AI developers are considered covert adversaries, who may attempt to circumvent guarantees if they can do so without detection [12]. These developers either have physical access to accelerators at data centers that they manage, or access them remotely through a cloud provider. Due to whistleblower risks, circumvention attempts would likely involve only a small number of trusted employees. If random inspections for tampering are conducted, any physical tampering would need to be either difficult to detect or limited to a small number of devices that could plausibly be reported as "lost" or otherwise unavailable for inspection.

Commercial entities might attempt to circumvent domestic regulations by:
1. Identifying or purchasing information about firmware vulnerabilities that offer plausible deniability if detected
2. Performing limited physical tampering on a strategic subset of devices
3. Exploiting ambiguities in guarantee logic specifications
4. Implementing modifications to cluster configurations that evade detection mechanisms

The Volkswagen emission scandal is one of the larger known examples of this type of deception. In the Volkswagen case, dozens of employees were involved in technical modifications that allowed vehicles to incorrectly pass emissions testing.

## International Treaty Verification

When flexHEG is deployed to verify compliance with international treaties, states themselves become potential adversaries. In this context, states are considered covert adversaries, as overt non-compliance would render verification systems unnecessary.

State actors possess sophisticated capabilities and advanced tools for circumvention. The scale of covert operations could be considerably larger than in domestic scenarios, potentially involving hundreds of personnel with specialized expertise, as this aligns with their core national security competencies.

A significant challenge in this scenario is that states subject to verification requirements may participate in the flexHEG design and manufacturing process. This participation creates opportunities to introduce subtle vulnerabilities, particularly in complex components that resist formal verification methods. For example, a state might influence secure enclosure designs to be vulnerable to specific types of laser attacks, or engineer interlock components with electrical coupling vulnerabilities that could be exploited under specific conditions. These types of vulnerabilities may be highly difficult to confidently protect against.

Even with these sophisticated offensive capabilities, certain circumvention methods would still be constrained by intelligence operations, randomized inspections, and robust tamper-evident packaging systems.

## International Treaty Enforcement

If FlexHEG is being used to enforce future adherence to certain rules, then the chip owner may be either a covert adversary as analyzed above, or an overt adversary in cases where a country decides to openly break their treaty commitments.

In the case of overt non-compliance, a state might deploy thousands of technical experts to identify exploitable security flaws. Upon discovering vulnerabilities in the flexHEG logic, a state could potentially develop scalable methods to exploit these vulnerabilities across their entire accelerator fleet. Similarly, if hardware security vulnerabilities are identified, the state could establish systematic processes to physically disable flexHEG mechanisms in multiple accelerators.

The feasibility of scaling hardware-based attacks depends on several factors: the success rate of hardware modifications, equipment availability, and the time and cost requirements of each modification. The required scale of the operation depends on the objective of the attack—exfiltrating sensitive data might require compromising only a single device, while circumventing licensing restrictions might require modifying thousands or millions of accelerators.

Beyond circumvention, there is also the risk that the flexHEG system itself could be used as an attack vector. In the international context, states might attempt to introduce backdoors or vulnerabilities that could enable sabotage of AI accelerators. This sabotage could manifest as overt disruption triggered by specific signals or timers, or as subtle manipulations designed to introduce unpredictable errors that impede development, similar to the approach used in the Stuxnet operation.

## Defensive Protection of AI Systems

FlexHEG can serve defensive purposes beyond regulation and treaty compliance. In these scenarios, the adversaries are external actors attempting to compromise AI infrastructure or exfiltrate sensitive intellectual property such as model weights or architecture details. For these defensive applications, the security challenge differs fundamentally—the goal is to prevent unauthorized access to protected accelerators and data rather than preventing circumvention of usage restrictions. Even compromising a single device could potentially enable the exfiltration of valuable model weights. Such attacks might target inference datacenters or be executed by rogue employees within training facilities.

## How Secure is Secure Enough?

It is not necessary for a flexHEG system to be absolutely secure against all possible attacks to serve its purpose effectively. Instead, the system should be secure enough that the cost of circumvention outweighs the benefits for most actors in most circumstances.

This approach is consistent with existing security paradigms in international relations. For example, the Biological Weapons Convention between the Soviet Union and United States did not achieve perfect verification, but raised the cost of covert violations through impromptu inspections. Similarly, flexHEG would aim to make rule violations sufficiently costly and risky to deter most potential violators.

The security of flexHEG could use two complementary strategies:

1. Technical measures: Tamper-evident enclosures, tamper-responsive self-disablement, cryptographic protocols for secure communication, and secure update mechanisms
2. External verification: Intelligence capabilities, randomized physical inspections, and whistleblower incentives

When evaluating whether flexHEG provides adequate security for a given application, policymakers should consider whether the system makes rule violations sufficiently difficult and detectable to alter the cost-benefit calculation for potential violators.

## Appendix C: Randomized Network Verification

The protocol below provides partial authenticity of network traffic at a small fraction of the cost of the full verification scheme ([Encrypted Cluster Formation](#)) using pseudo-randomized authenticity checks:

> 1. When a flexHEG device establishes a communication channel with another flexHEG device, it is privately sent (using public key cryptography[15]) a symmetric key and a random seed
> 2. The random seed is used to specify which subset of blocks in future transmissions will be authenticated (e.g. by pseudo-randomly sampling integers between 0:2N and skipping that many blocks)
> 3. All transmissions are sent with a "signature" for authenticity, but it is only sometimes real, depending on the randomization schedule
> 4. The FlexHEG devices that receive these messages randomly choose a smaller fraction that they will check, both to further reduce computational load and so that attackers cannot confidently use an extra flexHEG device to check if a block can be modified without detection.

Table 2: Pseudo-randomized authenticity checks (for reduced cryptographic load)

The pseudo-randomized authenticity checks system has several implementation details that would need to be considered:

Signatures would have to be slightly delayed because of limited capacity in the cryptographic hardware. The system could be configured to skip signatures of packets that are too close to previous ones, preventing the cryptographic hardware from becoming overwhelmed by bursts of traffic and maintaining a more consistent authentication rate[16].

---

[15] See [Encrypted Cluster Formation](#) for more on this network initialization process.
[16] Also, for more uniform processing, packets might need to be chunked to have close to uniform size.

Cryptography hardware would probably need to always be running to avoid detection of which packets are authenticated via side channels. If the cryptographic hardware only activates when authenticating specific blocks, an attacker might be able to detect the patterns of activity, revealing which blocks are being authenticated.

The system could implement semi-non-deterministic throttling if a discrepancy is detected, so an attacker couldn't easily use a 'canary' device to test if they can corrupt a packet.

By focusing only on selective authentication rather than full encryption of all traffic, the system can more easily coexist with other security mechanisms already in place. This randomized authentication protocol can use the same AES-GCM cryptography as previously discussed (though not using the cipher text), with N times greater efficiency. The downside of this protocol is that the network data is no longer confidential, and it could be modified en-route by an attacker with some probability (depending on N and the number of modifications) of avoiding detection. This uncertainty in authenticity makes the security of guarantees much more difficult to reason about and make robust. As a partial compromise, the protocol could be modified to mark some blocks as high-priority so that they are always authenticated (e.g. for exchanging information about the FLOP count or compute graph).